\documentclass[journal,comsoc]{IEEEtran}

\usepackage[dvips]{graphicx}
\usepackage{latexsym}
\usepackage{epsfig}
\usepackage{color}
\usepackage{amsmath} 
\usepackage{amssymb}
\usepackage{amsxtra}
\usepackage{enumitem}

\usepackage[T1]{fontenc}
\usepackage{setspace}

\usepackage{amsmath}
\usepackage[cmintegrals]{newtxmath}
\usepackage{graphicx}
\graphicspath{{Figures/}}
\usepackage{url}
\usepackage{cite}
\usepackage{algorithm}
\usepackage{algpseudocode}

\makeatletter
\def\BState{\State\hskip-\ALG@thistlm}
\makeatother

\usepackage{epstopdf}
\usepackage{amsmath}
\usepackage{amsxtra}
\usepackage{amstext}
\usepackage{amssymb}
\usepackage{latexsym}
\usepackage{dsfont} 
\usepackage{color}
\usepackage{multirow}
\usepackage{float}
\usepackage{hyperref}

\usepackage{tabularx,colortbl}
\usepackage[table]{xcolor}

\usepackage{lscape}

\usepackage{algorithm}
\usepackage{algpseudocode}

\usepackage{units}
\usepackage{cases}
\usepackage[ subrefformat=parens,labelformat=parens, caption=false, font=footnotesize]{subfig}
\usepackage{mathtools}
\mathtoolsset{showonlyrefs} 

\newcommand{\mbf}[1]{\mathbf{#1}}

\newcommand{\nth}[1]{{#1}{\text{th}}}

\newcommand{\abs}[1]{\left|{#1}\right|}
\newcommand{\norm}[1]{\left\|{#1}\right\|}

\DeclareMathOperator*{\argmax}{arg\,max}   
\DeclareMathOperator*{\argmin}{arg\,min}

\usepackage{mathtools}

\newcommand{\Hr}{\mathcal{H}}
\newcommand{\spp}{\mathrm{spp}}
\newcommand{\LoS}{\mathrm{LoS}}
\newcommand{\NLoS}{\mathrm{NLoS}}
\newcommand{\thh}{\mathrm{th}}
\newcommand{\Tx}{\mathrm{Tx}}
\newcommand{\GSM}{\mathrm{GSM}}
\newcommand{\GIM}{\mathrm{GIM}}
\newcommand{\opt}{\mathrm{opt}}

\begin{document}

\title{Terahertz-Band Ultra-Massive \\ Spatial Modulation MIMO}

\author{Hadi~Sarieddeen,~\IEEEmembership{Member,~IEEE,}
        Mohamed-Slim~Alouini,~\IEEEmembership{Fellow,~IEEE,}\\
        and~Tareq~Y.~Al-Naffouri,~\IEEEmembership{Senior Member,~IEEE}
\thanks{ Manuscript received December 2, 2018; revised April 27, 2019; accepted June 22, 2019. The work of all authors was supported by the King Abdullah University of Science and Technology (KAUST) research fund.

The authors are with the Department of Computer, Electrical and Mathematical Sciences and Engineering (CEMSE) at KAUST, Thuwal, Makkah Province, Saudi Arabia, 23955-6900 (e-mail: hadi.sarieddeen@kaust.edu.sa; slim.alouini@kaust.edu.sa; tareq.alnaffouri@kaust.edu.sa). }
}

\maketitle

\begin{abstract}

The prospect of ultra-massive multiple-input multiple-output (UM-MIMO) technology to combat the distance problem at the Terahertz (THz) band is considered. It is well-known that the very large available bandwidths at THz frequencies come at the cost of severe propagation losses and power limitations, which result in very short communication distances. Recently, graphene-based plasmonic nano-antenna arrays that can accommodate hundreds of antenna elements in a few millimeters have been proposed. While such arrays enable efficient beamforming that can increase the communication range, they fail to provide sufficient spatial degrees of freedom for spatial multiplexing. In this paper, we examine spatial modulation (SM) techniques that can leverage the properties of densely packed configurable arrays of subarrays of nano-antennas, to increase capacity and spectral efficiency, while maintaining acceptable beamforming performance. Depending on the communication distance and the frequency of operation, a specific SM configuration that ensures good channel conditions is recommended. We analyze the performance of the proposed schemes theoretically and numerically in terms of symbol and bit error rates, where significant gains are observed compared to conventional SM. We demonstrate that SM at very high frequencies is a feasible paradigm, and we motivate several extensions that can make THz-band SM a future research trend.

\end{abstract}


\maketitle

\section{Introduction}

\IEEEPARstart{T}{erahertz} (THz)-band communications \cite{kurner2014towards,Akyildiz6882305} are expected to play a pivotal role in the upcoming sixth generation (6G) of wireless mobile communications \cite{tariq2019speculative,saad2019vision}. Exploiting the large available bandwidths at the THz band between $\unit[0.03]{THz}$ and $\unit[10]{THz}$, unlike at the millimeter-wave (mmWave) band \cite{Rangan6732923} between 30 gigahertz (GHz) and $\unit[300]{GHz}$, has the potential to sustain terabit (Tb)/second data rate demands for many years to come. Since the THz band is conveniently situated between the microwave and optical bands, there has been growing interest among researchers on both sides of the spectrum to develop novel solutions for realizing THz communication infrastructures. For instance, one of the IEEE 802.15 wireless personal area networks (WPAN) study group's missions is to explore high-frequency ranges that support multi-gigabit (Gb)/second and Tb/second links. To this end, a THz interest group (IGthz) \cite{WPAN_THz} has been recently formed, and several experiments on THz wave propagation have been conducted. 

Due to the absence of compact THz signal sources and detectors that can operate at room temperature with high power and sensitivity (the so-called ``\emph{THz gap}''), THz-band applications have traditionally been in the areas of imaging and sensing \cite{1512493Woolard,4337845Liu}. Recent advancements in transceiver design, however, have allowed efficient THz signal generation, modulation and radiation. Hence, communication-based THz-band use cases are anticipated \cite{Zhang8663549}. In particular, THz communications can be used as alternatives to wired backbone connectivity in data centers \cite{celik2018wireless} or as part of large intelligent surface deployments \cite{nie2019intelligent}. The holy grail of THz communications, however, is enabling mobile wireless mid-range communications \cite{kurner2014towards,Akyildiz6882305,tariq2019speculative,saad2019vision}, in the context of device-to-device, drone-to-drone, vehicular, or personal communications. 


A variety of integrated electronic and photonic solutions for THz transceiver design have been proposed. Typical electronic solutions are based on silicon complementary metal-oxide-semiconductor (CMOS) technology \cite{Nikpaik8077757,Aghasi7819530}, and III-V-based semiconductors, such as heterojunction bipolar transistors (HBTs) \cite{Urteaga7915698}, high electron mobility transistors (HEMT) \cite{Deal8240460}, and Schottky diodes \cite{Mehdi7835091}. On the other hand, photonic solutions include uni-traveling carrier photodiodes \cite{Song6213156}, photoconductive antennas \cite{huang2017globally}, optical downconversion systems \cite{nagatsuma2016advances} and quantum cascade lasers \cite{lu2016room}. Furthermore, integrated hybrid electronic-photonic systems \cite{sengupta2018terahertz} have been proposed. In general, the target has shifted from designing perfect THz devices to designing efficient and programmable devices that are capable of satisfying emerging system-level properties.


However, plasmonic solutions, based on novel materials such as graphene \cite{hafez2018extremely,ju2011graphene,novoselov2012roadmap,ferrari2015science}, are currently being celebrated as the technology of choice at the THz band. Plasmonic materials possess superior electrical properties, such as very high electron mobility, and they support electron tunability and reconfigurability. In particular, plasmonic antennas support surface plasmon polariton (SPP) waves \cite{6708549Jornet,akyildiz2017graphene}, which have much smaller resonant wavelengths than free space waves, resulting in flexible and compact designs. Graphene-based sub-micrometric HEMT transistors can generate SPP waves, and a combination of plasmonic waveguides \cite{Leon6193116} and phase controllers \cite{Singh7481218} can deliver these signals to the desired antenna elements (AEs). Despite the fact that this approach is hindered by the low power of nano-transceivers\cite{knap2008plasma} and  short propagation length of SPP waves \cite{akyildiz2017graphene}, the small size of plasmonic sources allows them to be directly attached to each AE \cite{akyildiz2016realizing}, resulting in simplified efficient fully digital architectures.

Before THz communications can be realized, many more challenges need to be addressed from a communication system and signal processing perspective. For example, the very high propagation losses and power limitations at the THz-band result in very short communication distances. This problem can be mitigated by utilizing very dense nano-antenna arrays that provide high beamforming gains \cite{faisal2019ultra}. In particular, building on a similar proposal \cite{Torkildson6042312} for mmWave systems, a THz ultra-massive multiple-input multiple-output (UM-MIMO) solution that utilizes an array-of-subarrays (AoSA) of graphene-based nano-transceivers has been proposed \cite{akyildiz2017ultra,akyildiz2016realizing}. Similarly, frequency-dependent molecular absorptions result in band-splitting and bandwidth reduction at relatively large distances. To combat this problem, distance-aware and distance-adaptive solutions have been proposed, that optimize waveform design \cite{Han7321055}, resource allocation \cite{Han7490372} and beamforming \cite{Lin7116524}. Channel modeling at the THz-band is also a challenge due to the lack of realistic channel measurements (except for recently reported measurements up to $\unit[140]{GHz}$ \cite{xing2018propagation}). Nevertheless, ray-tracing-based channel models that assume the channel to be sparsely scattered and dominated by the line-of-sight (LoS) component have been proposed for graphene-based systems \cite{Jornet5995306,Han7579223,Han8387210,Han8417893}.  

The AoSA architecture \cite{akyildiz2017ultra,akyildiz2016realizing,Han8417893} provides the flexibility to trade beamforming with multiplexing, for better communication range or spectral efficiency. Also, plasmonic nano-antenna spacings can be significantly reduced while still avoiding mutual coupling effects \cite{Zakrajsek7928818}. In this paper, we address this trade-off between system performance and antenna design compactness. While compact designs are tempting, larger antenna separations result in better spatial diversity and spatial sampling, which enhances the performance of spatial multiplexing (SMX) and beamforming, respectively. One solution that could reap these benefits exploits the tunability property of plasmonics in a multicarrier design \cite{zakrajsek2017design}, where an interleaved antenna frequency map can maintain minimum spatial separations between same-frequency antennas while keeping all antenna elements (AEs) concurrently active. However, this is not a spectrum-efficient solution.

In this paper, we introduce and investigate the concept of spatial modulation (SM) \cite{Mesleh4382913} as a spectrum and power-efficient paradigm for THz UM-MIMO. To the best of our knowledge, SM at the THz-band has never been addressed in the literature. Hence, this article aims at motivating this paradigm as a future research trend. In fact, SM at very high frequencies is challenging because of LoS-dominance \cite{Liu7547944}. The contributions of this work can be summarized as follows:
\begin{enumerate}
  \item We analyze the channel condition at the THz-band as a function of frequency, communication range, and separations between antennas. Based on these parameters, favorable propagation settings that result in sufficient channel diversity are noted.
  \item We propose an adaptive hierarchical SM solution that maps information bits to antenna locations, at the level of subarrays (SAs) or AEs.
  \item We propose the use of a fully configurable graphene sheet, the dimensions of which can be adapted in real time for a target bit rate at a specific communication range.
  \item We provide an analytical closed-form symbol error rate (SER) performance analysis of the proposed schemes.
 \item We conduct extensive simulations to demonstrate the achievable gains of our proposed solutions. We illustrate that analytical and numerical results closely match.
  \item  We motivate several future research directions such as: SM with frequency-interleaved antenna maps, generic index modulation (IM), pulse-based THz SM, and generalized hierarchical detection and coding.
\end{enumerate}

The remainder of this paper is organized as follows: the system model is first presented in Sec.~\ref{sec:sysmodel}, followed by a detailed discussion on channel conditions at the THz-band in Sec.~\ref{sec:channel_cond}. The proposed SM solutions are then illustrated in Sec.~\ref{sec:propSM}, and the corresponding theoretical SER equations are derived in Sec.~\ref{sec:probSER}. Simulation results are presented in Sec.~\ref{sec:simulations}, possible extensions to the work are summarized in Sec.~\ref{sec:extensions}, and conclusions are drawn in Sec.~\ref{sec:conclusion}. Regarding notation, bold upper case, bold lower case, and lower case letters correspond to matrices, vectors, and scalars, respectively. Scalar norms, vector $\text{L}_2$ norms, and vector inner products are denoted by $\abs{\cdot}$, $\norm{\cdot}$, and $\langle\cdot,\!\cdot\rangle$, respectively. $\mathsf{E}[\cdot]$, $(\cdot)^{T}$, and $(\cdot)^{\Hr}$, stand for the expected value, transpose, and conjugate transpose, respectively. $\mathcal{N}(u,v)$ refers to a normal distribution of mean $u$ and variance $v$, and $Q(\cdot)$ refers to the Q-function, where $Q(x)\!=\!\int_x^{\infty} e^{-z^2/2}/\sqrt{2\pi}\ dz$. $\mbf{I}_N$ is an identity matrix of size $N$ and $j\!=\!\sqrt{-1}$ is the imaginary number. The system model notations are detailed in Tab. \ref{table:not}.

\begin{figure}[t]
\centering
\includegraphics[width=3in]{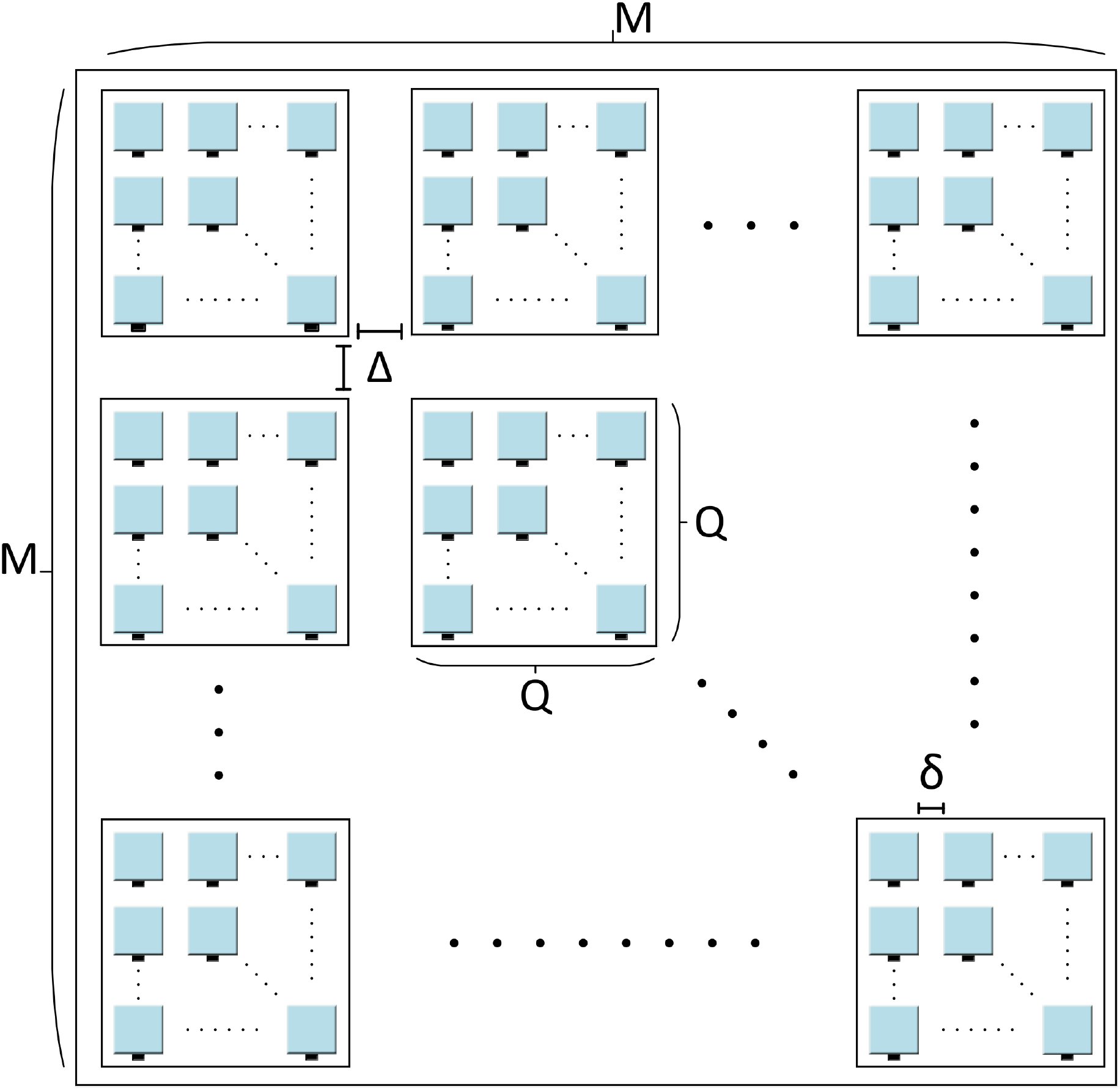}
\caption{Array-of-subarrayas antenna structure.}
\label{f:AoSAs}
\end{figure}

\section{System Model}
\label{sec:sysmodel}

We consider a three-dimensional UM-MIMO system model for graphene-based plasmonic nano-antennas as described in \cite{Han8417893}. The antennas consist of active graphene elements over a common metallic ground layer, with a dielectric layer in between. The AoSAs \cite{Lin7786122} at the transmitting and receiving sides consist of $M_t\!\times\!N_t$ and $M_r\!\times\!N_r$ SAs, respectively. Furthermore, we consider each SA to consist of a $Q\times Q$ set of nano-AEs. The resultant configuration can be represented as a \emph{large} $M_tN_tQ^2\times M_rN_rQ^2$ MIMO system \cite{Sarieddeen8186206} by vectorizing the two-dimensional antenna indices on each side. Note that this configuration differs from conventional asymmetric \emph{massive} MIMO, with large antenna arrays at a transmitting base station, and typically single-antenna users at the receiving side. We denote by $\delta$ and $\Delta$ the distances between two adjacent nano-AEs and two adjacent SAs, respectively, at both the transmitter and the receiver sides, without loss of generality.  

We assume single-carrier LoS transmission with frequency flat fading. The end-to-end baseband system model at a specific frequency can be expressed as
\begin{equation}\label{eq:sysmodel}
	\mbf{y} = \mbf{W}^{\Hr}_r \mbf{H}\mbf{W}^{\Hr}_t \mbf{x} + \mbf{W}^{\Hr}_r \mbf{n},
\end{equation} 
where $\mbf{x}\!=\![x_{1}x_{2}\cdots x_{N_s}^{}]^{T}\!\in\!\mathcal{X}^{N_s\times1}$ is the information-bearing modulated symbol vector, with symbols belonging to the same quadrature amplitude modulation (QAM) constellation $\mathcal{X}$, $\mbf{y}\!\in\!\mathcal{C}^{N_s\times1}$ is the received symbol vector,  $\mbf{H}\!=\![\mbf{h}_{1}\mbf{h}_{2}\cdots \mbf{h}_{M_tN_t^{}}]\!\in\!\mathcal{C}^{M_rN_r\times M_tN_t}$ is the channel matrix, $\mbf{W_t}\!\in\!\mathcal{R}^{N_s \times M_tN_t}$ and $\mbf{W_r}\!\in\!\mathcal{R}^{M_rN_r\times N_s}$ are the baseband precoder and combiner matrices, which define  SA utilization, and $\mbf{n}\!\in\!\mathcal{C}^{M_rN_r\times1}$ is the additive white Gaussian noise (AWGN) vector of power $\sigma^2$.

\begin{table}[t]
\small
\caption{Summary of Important Abbreviations}\label{table:abre}
\centering
 \begin{tabular}{| c || c |} 
 \hline
 Parameter & Value  \\ 
 \hline\hline
AE & antenna element \\ \hline
AoSA & array-of-subarrays \\ \hline
AWGN & additive white Gaussian noise \\ \hline
BER & bit error rate \\ \hline
GIM & generalized index modulation \\ \hline
GSM & generalized spatial modulation \\ \hline
HITRAN & high resolution transmission database \\ \hline
IM & index modulation \\ \hline
LoS & line-of-sight \\ \hline
MIMO & multiple-input multiple-output \\ \hline
mmWave & millimeter-wave \\ \hline
MRRC & maximum receive ratio combining \\ \hline
NLoS & non-line-of-sight \\ \hline
QAM & quadrature amplitude modulation \\ \hline
RF & radio frequency \\ \hline
SA & subarray \\ \hline
SER & symbol error rate \\ \hline
SM & spatial modulation \\ \hline
SMX & spatial multiplexing \\ \hline
SNR & signal-to-noise-ratio \\ \hline
SPP & surface plasmon polariton \\ \hline
UM-MIMO & ultra-massive MIMO \\ \hline

 \hline

 \end{tabular}
\end{table}

\begin{table}[h]
\small
\caption{Summary of System Model Notations}\label{table:not}
\centering
 \begin{tabular}{| c || c |} 
 \hline
Parameter & Value  \\ 
 \hline\hline

$\mbf{a}_t,\mbf{a}_r$ & transmit and receive SA steering vectors \\ \hline
$\mbf{b}$ & binary symbol vector \\ \hline
$\mbf{C}_t,\mbf{C}_r$ & transmit and receive mutual coupling matrices \\ \hline
$D$ & communication range \\ \hline
$d$ & effective communication distance \\ \hline
$f$ & carrier frequency \\ \hline
$G_t,G_r$ & transmit and receive antenna gains \\ \hline
$\mbf{g}$ & transformed received symbol vector \\ \hline
$\mbf{H}$ & channel matrix \\ \hline
$\mathcal{K}(f)$ & absorption coefficient \\ \hline
$M_r,N_r$ & SA indices in receiving AoSA  \\ \hline
$M_t,N_t$ & SA indices in transmitting AoSA  \\ \hline
$N_b$ & number of information bits \\ \hline
$\mbf{n}$ & noise vector \\ \hline
$Q$ & AE index in a SA ($Q\times Q$ AE's in a SA) \\ \hline
$\mbf{W}_t,\mbf{W}_r$ & precoding and combining matrices \\ \hline
$\mathcal{X}$ & modulation constellation \\ \hline
$\mbf{x}$ & transmitted symbol vector \\ \hline
$\mbf{y}$ & received symbol vector \\ \hline

$\alpha$ & path gain \\ \hline
$\gamma$ & signal-to-noise ratio \\ \hline
$\Delta$ & distance between two adjacent SAs \\ \hline
$\bar{\Delta}$ & quantized optimal $\Delta$ \\ \hline
$\delta$ & distance between two adjacent AEs \\ \hline
$\lambda$ & free-space wavelength \\ \hline
$\lambda_{\spp}$ & SPP wavelength \\ \hline
$\sigma^2$ & noise power \\ \hline
$\Phi$ & AE phase shifts \\ \hline
$\phi_t$,$\theta_t$ & transmit angles of departure and arrival \\ \hline
$\phi_r$,$\theta_r$ & receive angles of departure and arrival \\ \hline
$\psi$ & AE coordinate positions \\ \hline

 \hline

 \end{tabular}
\end{table}

Each SA is assumed to generate a single beam due to beamforming at the level of AEs. An element of $\mbf{H}$, $h_{m_rn_r,m_tn_t}$, the frequency response between the $(m_t,n_t)$ and $(m_r,n_r)$ SAs, is thus defined as 
\begin{equation}\label{eq:channel}
	h_{m_rn_r,m_tn_t} = \mbf{a}^{\Hr}_r(\phi_r,\theta_r)G_r \alpha_{m_rn_r,m_tn_t} G_t \mbf{a}_t(\phi_t,\theta_t),
\end{equation} 
for $m_r\!=\!1,\cdots,M_r$, $n_r\!=\!1,\cdots,N_r$, $m_t\!=\!1,\cdots,M_t$, and $n_t\!=\!1,\cdots,N_t$, where $\alpha$ is the path gain, $\mbf{a}_t$ and $\mbf{a}_r$ are the transmit and receive SA steering vectors, $G_t$ and $G_r$ are the transmit and receive antenna gains of the Friis formula \cite{842121Andersen}, and $\phi_t$,$\theta_t$ and $\phi_r$,$\theta_r$ are the transmit and receive angles of departure and arrival, respectively, with $\phi$'s being the azimuth angles and $\theta$'s the elevation angles. The steering vectors can be expressed in terms of $\mbf{C}_t,\mbf{C}_r\!\in\!\mathcal{R}^{Q^2\times Q^2}$, the mutual coupling matrices of transmit and receive graphene-based arrays, as $\mbf{a}_t(\phi_t,\theta_t)\!=\!\mbf{C}_t\mbf{a}_0(\phi_t,\theta_t)$ and $\mbf{a}_r(\phi_r,\theta_r)\!=\!\mbf{C}_r\mbf{a}_0(\phi_r,\theta_r)$. Note that we neglect the effect of mutual coupling in the remainder of this work ($\mbf{C}_t\!=\!\mbf{C}_r\!=\!\mbf{I}_{Q^2}$) by assuming $\delta\!\geq\!\lambda_{\spp}$ \cite{Zakrajsek7928818}, where $\lambda_{\spp}$ is the SPP wavelength ($\lambda_{\spp}$ is much smaller than $\lambda$, the free-space wavelength). The ideal transmit SA steering vector is defined as
\begin{equation}\label{eq:steering}
	\mbf{a}_0(\phi_t,\theta_t) \!=\! \frac{1}{Q} [e^{j\Phi_{1,1}},\!\cdots\!,e^{j\Phi_{1,Q}},e^{j\Phi_{2,1}},\!\cdots\!,e^{j\Phi_{p,q}},\!\cdots\!,e^{j\Phi_{Q,Q}}]^T,
\end{equation} 
where the phase shift corresponding to AE $(p,q)$ is
\begin{multline}\label{eq:shifts}
	\Phi_{p,q} = \psi_x^{(p,q)}\frac{2\pi}{\lambda_{\spp}}\cos \phi_t \sin \theta_t \\ + \psi_y^{(p,q)}\frac{2\pi}{\lambda_{\spp}}\sin \phi_t \sin \theta_t  + \psi_z^{(p,q)}\frac{2\pi}{\lambda_{\spp}}\cos \theta_t,
\end{multline} 
with $\psi_x^{(p,q)}$, $\psi_y^{(p,q)}$, and $\psi_z^{(p,q)}$ being the coordinate positions of AEs in the three-dimensional space. At the receiver side, $\mbf{a}_0(\phi_r,\theta_r) $ can be similarly defined. For simplicity and without loss of generality, we assume symmetry in the remainder of this work, i.e., we assume $M_t \!=\! M_r \!=\! N_t \!=\! N_r \!=\! M$. The resultant AoSA is illustrated in Fig.~\ref{f:AoSAs}. Assuming normalized symbols ($\mathsf{E}[\mbf{x}^{\mathcal{H}}\mbf{x}^{}]\!=\!1$) and perfect beamforming-angle alignment, the signal-to-noise-ratio (SNR) per stream can be expressed as 
\begin{equation}\label{eq:SNR}
	\gamma = G_tG_rQ^2\abs{\alpha}^2/\sigma^{2}.
\end{equation} 
Beamfoarming makes the SA the basic and smallest addressable component of this model. However, nano-AEs can still be individually addressed in a fully digital design when $Q\!=\!1$, but the resultant architecture is much more complex.

\section{Channel Condition at the THz-Band}
\label{sec:channel_cond}

Signal propagation at THz frequencies is quasi-optical. Due to large reflection losses, the channel is dominated by the LoS path, and possibly very few, if any, non-LoS (NLoS) reflected rays. Scattered and refracted rays can be neglected. The frequency-dependent LoS path gain is
\begin{align}\label{eq:LoS}
	\alpha_{m_rn_r,m_tn_t}^{\LoS} =  & \frac{c}{4\pi f d_{m_rn_r,m_tn_t}} \\ & \times e^{ -\frac{1}{2} \mathcal{K}(f) d_{m_rn_r,m_tn_t} }  e^{  -j \frac{2\pi f}{c} d_{m_rn_r,m_tn_t}},
\end{align} 
where $f$ is the frequency of operation, $c$ is the speed of light in vacuum, $d_{m_rn_r,m_tn_t}$ is the distance between the transmitting and receiving SAs, and $\mathcal{K}(f)$ is the absorption coefficient \cite{Jornet5995306}. Due to the beamforming gain, a single dominant ray exists between the transmitting and receiving SAs. We hereby assume this to be the LoS path. 

Supporting simultaneous transmission of multiple data streams at lower frequencies, up to a few GHz, can be easily achieved. This is facilitated by rich scattering at such frequencies which results in high-rank channel matrices. However, a LoS channel at low frequencies can only possess a single spatial degree of freedom. Nevertheless, achieving good multiplexing gains in strong LoS environments at very high frequencies is feasible. The latter is true when antenna spacings are much larger than the operating wavelength, which is a manageable design constraint. The optimal (largest) number of spatial degrees of freedom (DoF) in a LoS environment at high frequencies is achieved by sparse antenna arrays \cite{Torkildson6042312} that result in sparse multipath environments. 

As per our notation, these arrays can be constructed by tuning the SA spacing $\Delta$, such that orthogonal eigenmodes are produced in SMX, which generates spatially uncorrelated channel matrices.  The optimal $\Delta$ is a function of $\lambda$ and $d$, where shorter $\lambda$ and smaller $d$ both lead to a smaller optimal antenna separation. For SM, similar high rank channels are required. The relation between channel condition and design compactness in a SM setup was previously formulated as an optimization problem at mmWave \cite{Liu7547944}, in the context of optimal maximum likelihood (ML) detection. We hereby analyze this trade-off at the THz-band.

\begin{figure}[t]
\centering
\includegraphics[width=3.4in]{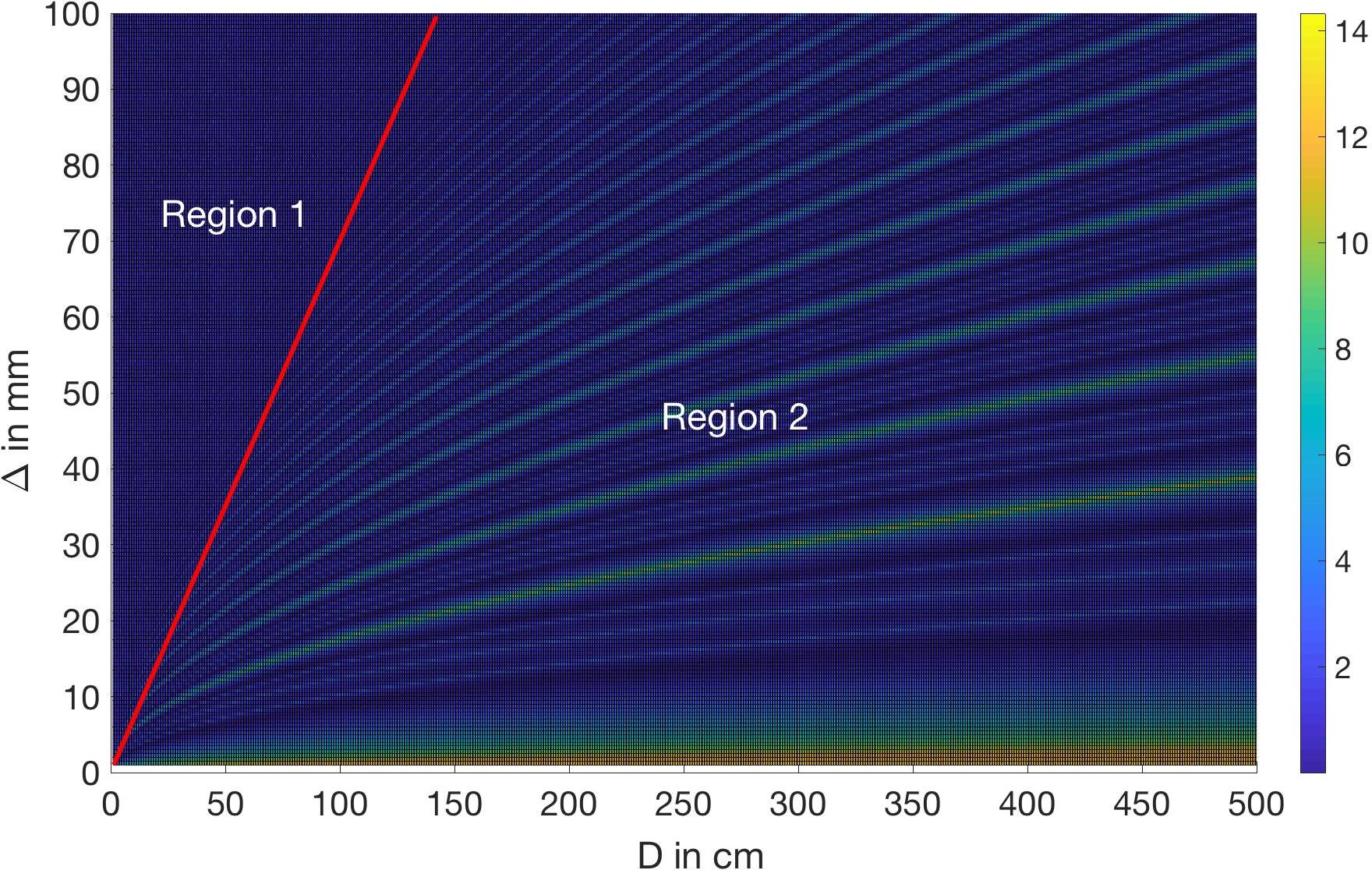}
\caption{Channel condition number as a function of $\Delta$ and $D$ at $\unit[1]{THz}$.}
\label{f:channel_2}
\end{figure}

\begin{figure}[t]
\centering
\includegraphics[width=3.5in]{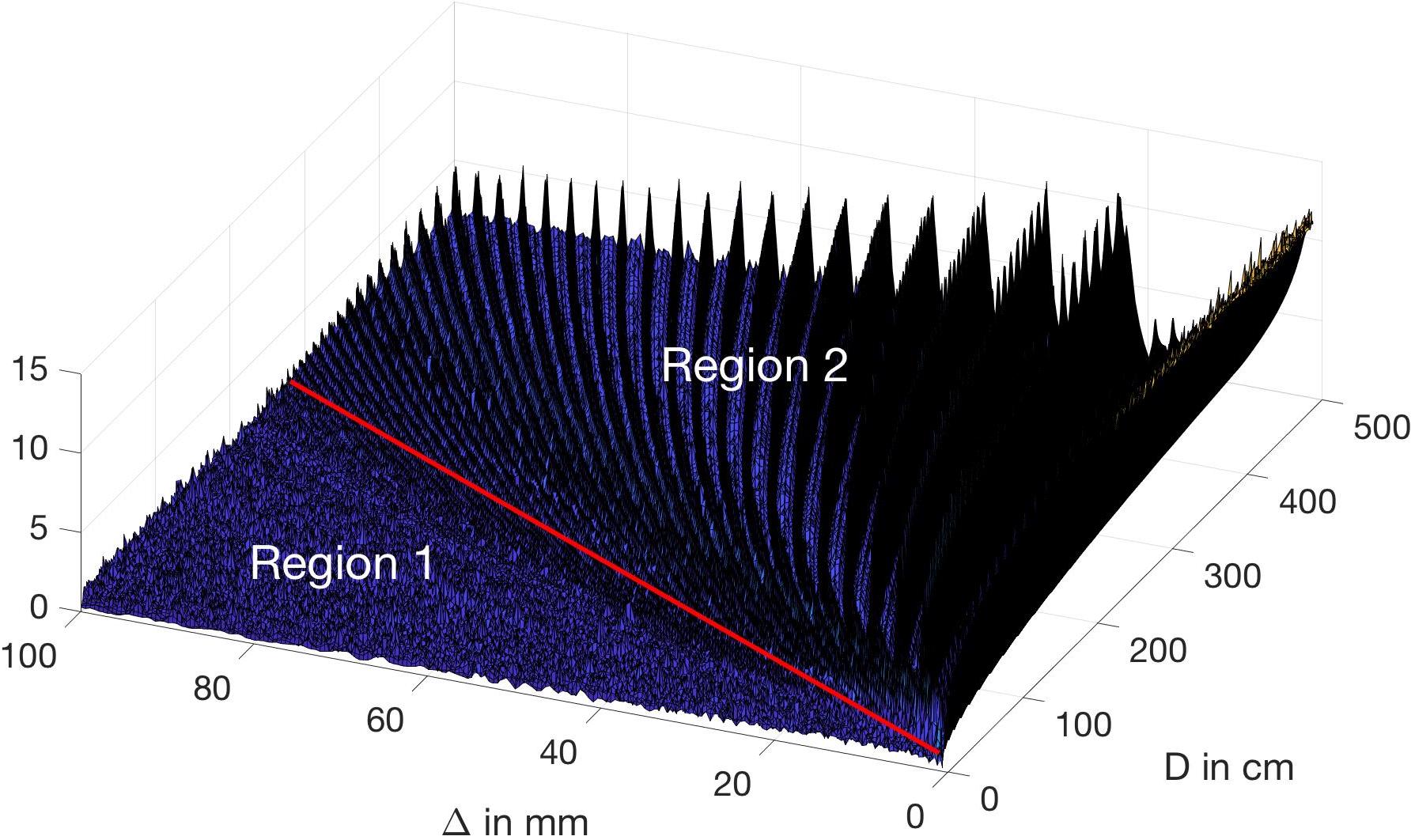}
\caption{Channel condition number as a function of $\Delta$ and $D$ at $\unit[3]{THz}$.}
\label{f:channel_3}
\end{figure}

For uniformly spaced AoSAs, the effective distance $d_{m_rn_r,m_tn_t}$ can be expressed in terms of the communication range $D$, which is the distance between the centers of transmitting and receiving antenna arrays, as $d_{m_rn_r,m_tn_t}\!=\!\sqrt{D^2 \!+\! \Delta^2\left( (m_r\!-\!m_t)^2 + (n_r\!-\!n_t)^2 \right)}$. Furthermore, for $D\!\gg\!\Delta$, we have via binomial approximation $d_{m_rn_r,m_tn_t}\!\approx\!D \!+\! \Delta^2\left( (m_r\!-\!m_t)^2 + (n_r\!-\!n_t)^2 \right)\!/2D$. Hence, the channel gain can be approximated as
\begin{equation}\label{eq:LoS_approx}
	\alpha_{m_rn_r,m_tn_t}^{\LoS} \approx \frac{c}{4\pi f D} e^{ -\frac{1}{2} \mathcal{K}(f) D } e^{  -j \frac{2\pi f}{c} \left( D + \frac{\Delta^2\left( (m_r-m_t)^2 + (n_r-n_t)^2 \right)}{2D}\right)}. 
\end{equation}
The capacity of our $M^2\!\times \!M^2$ MIMO system is maximized when all $M$ singular values of $\mbf{H}$ are equal, which is the case when all columns of $\mbf{H}$ are orthogonal. Let $(m_t\!=\!k,n_t\!=\!l)$ and $(m_t\!=\!\acute{k},n_t\!=\!\acute{l})$ be the coordinates of two arbitrary transmitting SAs. Neglecting antenna gains, the inner product between the corresponding channel columns is
\begin{align}\label{eq:orthogonality}
	\langle \mbf{h}_{kl},\mbf{h}_{\acute{k}\acute{l}}  \rangle =& \left(\frac{c}{4\pi f D}\right)^2 e^{ -\mathcal{K}(f) D } \\ & \times \sum_{m_r=0}^{M-1}\sum_{n_r=0}^{M-1}  e^{  j \frac{\pi f}{cD} \Delta^2\left( (m_r\!-\!k)^2 \!+\! (n_r\!-\!l)^2 \!-\! (m_r\!-\!\acute{k})^2 \!-\! (n_r\!-\!\acute{l})^2 \right)}\\
	=& \left(\frac{c}{4\pi f D}\right)^2 e^{ -\mathcal{K}(f) D } \\ & \times \sum_{m_r=0}^{M-1}  e^{  j \frac{2\pi f}{cD} \Delta^2 m_r\left(\acute{k}-k \right)} \sum_{n_r=0}^{M-1}  e^{  j \frac{2\pi f}{cD} \Delta^2 n_r\left(\acute{l}-l \right)} \\
  \propto & \frac{ \sin \left( \pi M (\acute{k}-k) \frac{\Delta^2 f}{cD}\right) }{\sin \left( \pi (\acute{k}-k) \frac{\Delta^2 f}{cD}\right)}\frac{ \sin \left( \pi M (\acute{l}-l) \frac{\Delta^2 f}{cD}\right) }{\sin \left( \pi (\acute{l}-l) \frac{\Delta^2 f}{cD}\right)}.
\end{align} 
The channels are orthogonal when this product is zero, i.e., for an optimal value of $\Delta$
\begin{equation}\label{eq:innerprod}
	\Delta_{\opt} = \sqrt{z\frac{Dc}{Mf}},
\end{equation} 
for integer values of $z$. Hence, condition \eqref{eq:innerprod} guarantees optimal AoSA design. Note that the same condition was observed in \cite{Larsson1543276}, for uniform rectangular arrays in low-frequency LoS systems, by maximizing the channel matrix determinant as a means to maximize capacity.

The channel condition number \cite{Klema1102314}, which is the ratio of the largest to the smallest singular value of $\mbf{H}$, is illustrated in figures \ref{f:channel_2} and \ref{f:channel_3}, at $\unit[1]{THz}$ and $\unit[3]{THz}$, respectively, as a function of $\Delta$ and $D$. The smaller the singular value is, the better conditioned the channels are (preferably $\!<\!\unit[10]{dB}$). Two regions of operation can be seen (separated by a red line). In \emph{Region 1}, $\Delta$ is sufficiently large compared to $D$ (relatively short communication ranges), which makes different paths between transmitting and receiving SAs sufficiently distinct, and hence the channel is always well-conditioned. In \emph{Region 2}, however, for relatively large communication ranges, the channel paths become highly correlated. The low-value curves (dark blue curves in Fig. \ref{f:channel_2} and surf dips in Fig. \ref{f:channel_3}) are simply plots of \eqref{eq:innerprod} for different values of $z$, which correspond to the cases when all singular values of $\mbf{H}$ are equal (the curves are denser at higher frequencies). At these locations, the singular values correspond to eigenchannels over which multiple data streams can be transmitted (in SMX) or modulated (SM).

 \begin{figure}[t]
\centering
\includegraphics[width=3in]{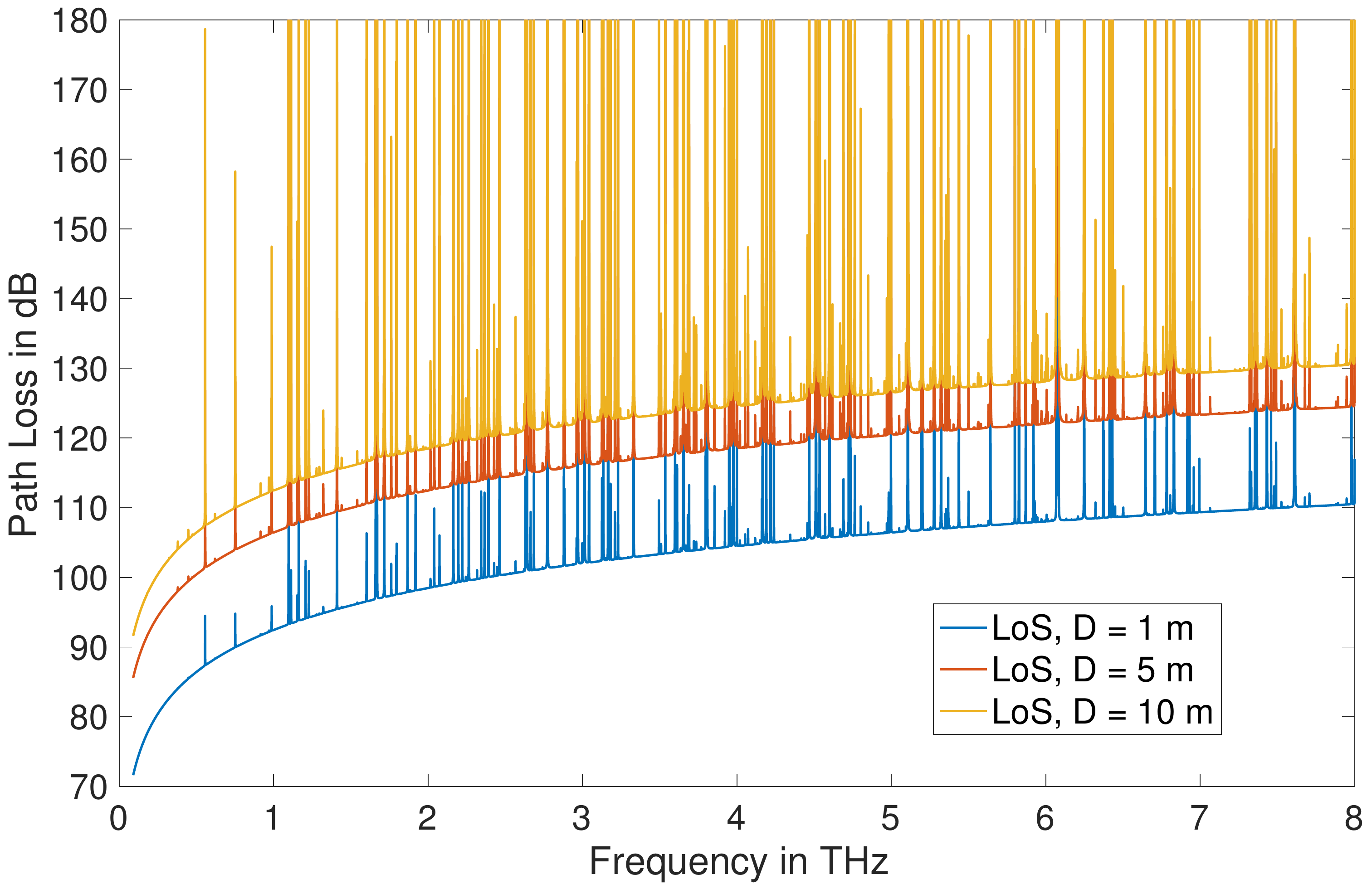}
\caption{Path loss as a function of frequency for different  ranges.}
\label{f:channel_1}
\end{figure}

The path loss is then illustrated in Fig. \ref{f:channel_1}, versus frequency over the THz band, for different communication ranges. The path loss is dominated by spikes resulting from molecular absorptions, mainly due to water vapor. The equations that describe the attenuations due to such absorptions are detailed in \cite{Jornet5995306}, where all required data can be extracted from the high resolution transmission (HITRAN) \cite{gordon2017hitran2016} database. Note that for distances beyond $\unit[1]{m}$ and frequencies higher than $\unit[1]{THz}$, the available spectrum shrinks and splits into multiple smaller sub-bands (in which channels behave as flat). The achievable bandwidth is thus distance dependent.

However, this is not the only effect that molecular absorption has on the system. In fact, such absorptions are typically followed by re-radiations with negligible frequency shifts. This phenomenon is assumed to result in favorable rich scattering in \cite{Hoseini8269042}. Nevertheless, a more realistic model lumps the effect of such coherent re-radiations in a molecular absorption noise factor \cite{Jornet5995306}. Consequently, and since graphene-based electronic devices are low-noise in nature \cite{pal2009ultralow}, $\mbf{n}$ is dominated by the channel-induced component at spike locations, and the resultant noise is colored over frequency. Unless explicitly stated, all analysis and numerical results in the remainder of this work assume operations in absorption-free spectrum windows, i.e., $ \mathcal{K}(f)\!=\!0$ and the noise is $\mathcal{N}(0,\sigma^2)$.

\section{Proposed SM Schemes}
\label{sec:propSM}

Depending on inter-antenna separations and the communication range, we define two modes of operation. The first mode corresponds to operations in \emph{Region 1}. Here, we propose an adaptive two-level SM solution, in which antenna allocations are made at the level of SAs or AEs. In the second mode of operation (\emph{Region 2}), we have $D\!\gg\!\Delta$, where the yellow peak curves of Fig. \ref{f:channel_2} must be avoided. For this mode, we propose an adaptive design based on a configurable graphene sheet that dynamically adjusts inter-antenna separations to satisfy condition \eqref{eq:innerprod}. Note that in both modes, physical AE separations ($\delta$) and dimensions are assumed to be fixed, as dictated by the target operational frequency of a specific chip. A frequency adaptive design is later motivated in Sec. \ref{sec:interleaved_map}.

\subsection{Adaptive Hierarchical SM}
\label{sec:dual_layer}

\begin{figure}[t]
\centering
\includegraphics[width=2.7in]{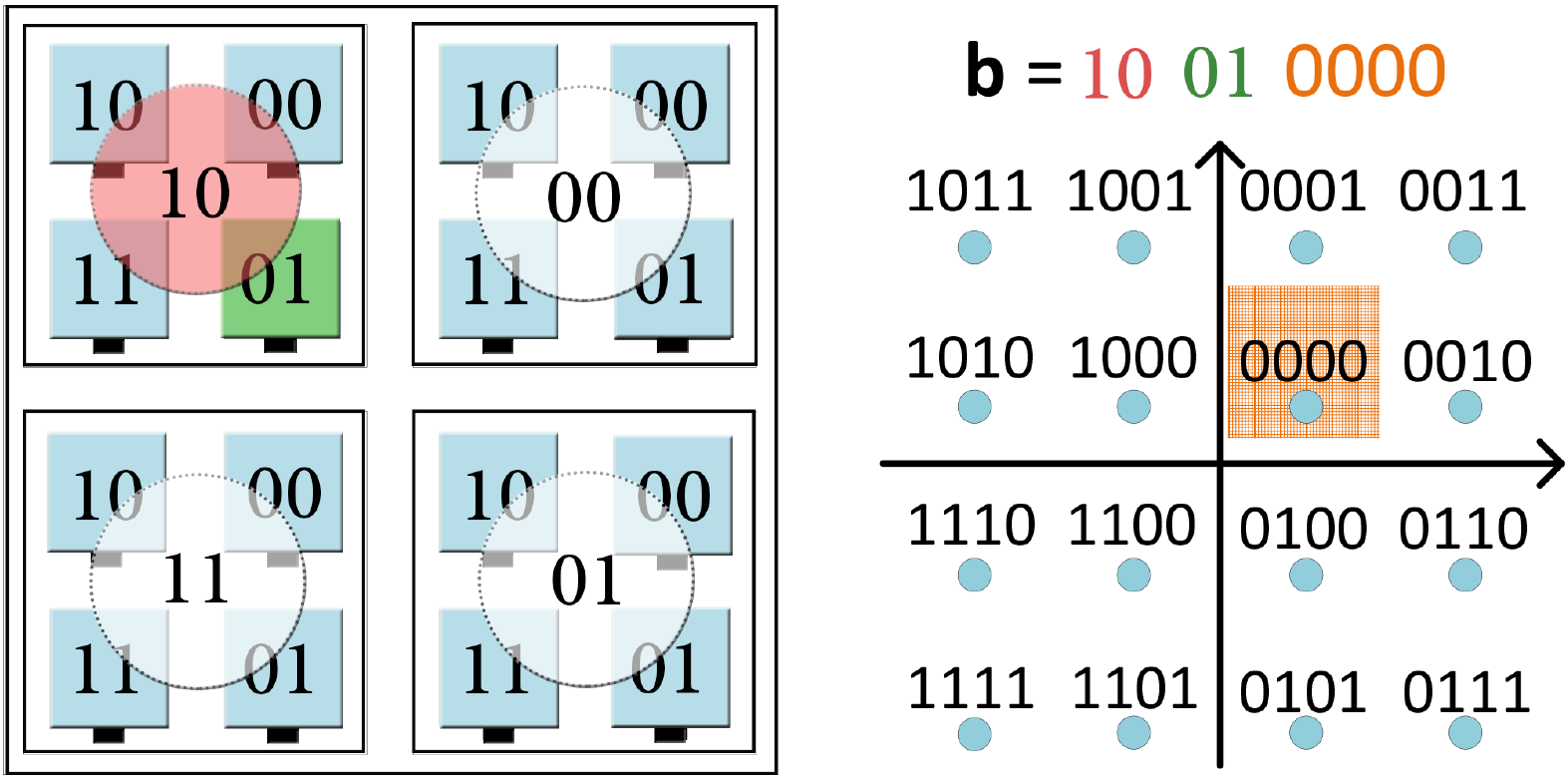}
\caption{Hierarchical modulation: $M=2$, $Q=2$, and $\mathcal{X}$ is 16-QAM.}
\label{f:mapper}
\end{figure}

For the first mode of operation, we slightly change the notation of Sec. \ref{sec:sysmodel} by assuming SM instead of beamforming at the level of AEs. This is well-founded for two reasons. First, short distance communications ($D\!\leq\!\unit[1]{m}$) do not require beamforming gains to combat path loss. Second, $\delta\!<\!\Delta$ can also be sufficiently large for a given $D$, resulting in well-conditioned channels for SM at the AE level. We propose a generic hierarchical SM solution tailored for the peculiarities of the corresponding architecture. The total number of bits that can be transmitted in a single channel use is
\begin{equation}\label{eq:nbofbits}
	N_b =  \underbrace{\log_2\left( M^2 \right)}_\text{SA} + \underbrace{\log_2\left( Q^2 \right)}_\text{AE} + \underbrace{\log_2\left( \abs{\mathcal{X}} \right)}_\text{symbols}.
\end{equation}  
Let $\mbf{b}\!=\![\mbf{b}_m \mbf{b}_q \mbf{b}_s]\!\in\!\{0,1\}^{N_b}$ be the binary vector that is transmitted over one symbol duration. We have: $\mbf{b}_m\!\in\!\{0,1\}^{\log_2\left( M^2 \right)}$ is the bit representation of the selected SA, $\mbf{b}_q\!\in\!\{0,1\}^{\log_2\left( Q^2 \right)}$ is that of the selected AE, and $\mbf{b}_s\!\in\!\{0,1\}^{\log_2\left( \abs{\mathcal{X}} \right)}$ is that of the QAM symbol. Hence, the number of SAs, number of AEs, and constellation size can be traded for a target bit rate. While $\Delta$ is assumed to always satisfy first mode conditions, we monitor $\delta$ to decide whether SM is to be carried at the level of AE or SA. For example, if $D\!<\!20\delta$ at $\unit[1]{THz}$ (roughly speaking based on the red line in Fig. \ref{f:channel_2}), hierarchical SM holds, and the corresponding channel dimension is $M_tN_tQ^2\times M_rN_rQ^2$. Otherwise, SM is retained at the level of SAs, where only the first antenna element in each SA can be activated. In the latter case, the channel dimension is $M_tN_t\times M_rN_r$, and we have $\mbf{b}\!=\![\mbf{b}_m  \mbf{b}_s]$. This adaptive design can be aided by real-time feedback of a performance metric such as SER, or by simple decision thresholds that can be fetched from a training-based lookup table that is indexed by $D$ and $f$. Selecting a symbol from the QAM constellation $\mathcal{X}$ based on the components of $\mbf{b}_s$ follows a Gray mapping criterion, similar to that used in the Long-Term Evolution (LTE) standard \cite{LTE_36.211}, for example. We propose using the same mapper to map $\mbf{b}_m$ and $\mbf{b}_q$ to antenna locations, in a layered manner \cite{Hemadeh8281511,Meric6678127}, as shown in Fig. \ref{f:mapper}.

At the receiver side, we assume perfect channel knowledge. The active AE (assuming SM at the level of AE) is estimated as $\hat{l}$ via maximum receive ratio combining (MRRC). The received vector first gets multiplied by the Hermitian conjugate of the channel matrix. Since the components of $\mbf{x}$ are all zeroes except at the location of the active AE, $\hat{l}$ can be estimated as the index of the transformed received vector element that has the maximum magnitude. Hence, we have
\begin{align}\label{eq:MRRC}
	\mbf{g} &= \mbf{H}^{\mathcal{H}}\mbf{y} \\
	\hat{l} &= \argmax_l \left( \abs{g_l} \right) \ \ \ \ l=1,\cdots,M^2Q^2,
\end{align} 
where $g_l$ is the $\nth{l}$ element of $\mbf{g}$. The detected transmitted symbol $\hat{x}_{\hat{l}}$, the $\nth{\hat{l}}$ component of the detected vector $\hat{\mbf{x}}$, can then be obtained as 
\begin{equation}\label{eq:sym_det}
	\hat{x}_{\hat{l}} = \left\lfloor \frac{1}{\mbf{h}_{\hat{l}}^{\mathcal{H}}\mbf{h}_{\hat{l}}} g_{\hat{l}} \right\rceil_{\mathcal{X}}
\end{equation} 
where $\lfloor \beta \rceil_{\mathcal{X}} \triangleq \argmin_{x \in \mathcal{X}} \abs{\beta-x}$ is the slicing operator on constellation $\mathcal{X}$. Finally, the SM demodulator performs bit demapping to recover the transmitted information.

\subsection{Configurable Graphene Sheet}
\label{sec:graphene_sheet}

The fact that SM increases spectral efficiency by only the base-two logarithm of the number of antennas is understood to be a disadvantage compared to SMX, which increases spectral efficiency by a multiplicative factor equal to the number of transmit antennas. However, since the number of graphene-based AEs can be made very large at low costs and footprints, system efficiency can be retained with SM at the THz-band. To this end, we propose an architecture that consists of two large antenna arrays, one at the transmitter side and another at the receiver side, that can be reprogrammed in real time depending on system parameters. Such arrays are in effect two graphene sheets consisting of a sufficiently large number of antenna elements, that are uniformly separated by a distance $\delta$. We assume, for simplicity, that these two sheets are matched  (symmetric large MIMO). Hence, we limit the discussion to calculating transmit array parameters. 

For the second mode of operation (\emph{Region 2}), we consider a setting where the number, size, and corresponding locations of SAs are configurable. In the first configuration step, the number of AEs in a SA is determined. Depending on the frequency of operation, a specific power gain is required to combat the path loss at a specific communication distance. This can be provided by a combination of antenna gains and array gains \eqref{eq:steering}. In particular, a path loss threshold $PL_{\thh}$ is defined based on the link budget \cite{Schneider6155633,Han7490372} as
\begin{multline}\label{eq:link_budget}
	PL_{\thh} = P_{\Tx} + 10\log_{10}(G_t) + 10\log_{10}(G_r) \\ + 20\log_{10}(\abs{\alpha})  - \gamma_{\thh} - 20\log_{10}(\sigma),
\end{multline} 
where $P_{\Tx}\!=\!\unit[10]{dBm}$ is the transmit power, $\gamma_{\thh}\!=\!\unit[10]{dB}$ is the threshold SNR, and $20\log_{10}(\sigma)\!=\!\unit[-80]{dBm}$ (assuming absorption-free spectral windows). As shown in Fig. \ref{f:channel_1}, for frequencies less than $\unit[1]{THz}$, $PL_{\thh}$ is $\unit[90]{dB}$ for $D\!=\!\unit[1]{m}$, and $\unit[110]{dB}$ for $D\!=\!\unit[10]{m}$. Hence, we should have $10\log_{10}(G_t) \!+\! 10\log_{10}(G_r)\!+\!20\log_{10}(\abs{\alpha})$ equal to $\unit[10]{dBi}$ in the former case, and $\unit[20]{dBi}$ in the latter, where the term $20\log_{10}(\abs{\alpha})$ embeds the array gain. Based on the required array gain, the number of AEs Q for beamforming is determined. For example, under perfect beamforming angle synchronization, a $64$-element SA with $Q\!=\!8$ can provide an $\unit[18]{dB}$ array gain.

The second configuration step consists of finding the optimal antenna spacing. Note that $\delta\!=\!\lambda/2$ can be retained as the best (maximum acceptable) AE separation within a SA, which maximizes the beamforming gain while avoiding grating lobe effects \cite{hansen2009phased}. However, $\Delta$ should be adapted for each $D$ as per \eqref{eq:innerprod}. A target $\Delta_{\opt}$ is thus achieved with a resolution of $\pm\delta/2$ by activating subarrays that are $\bar{\Delta}$ apart, where
\begin{equation}\label{eq:delta_prime}
	\bar{\Delta} = \left\lfloor \frac{1}{\delta}\sqrt{z\frac{Dc}{Mf}} \right\rceil_{\mathcal{Z}} \delta.
\end{equation} 
Note that at the receiver side, the operations of equations \eqref{eq:MRRC} and \eqref{eq:sym_det} hold for this design, but for values $l=1,\cdots,M^2$, with $M^2$ being the number of active SAs in the sheet, since beamforming is always assumed within SAs.

The key for such flexibility in the design is the ability to control a specific subset of AEs, whether it is contiguous or not, and this is directly linked to the excitation mechanism. SPP wave excitation can be achieved through optical or electrical pumping. Optical pumping \cite{williams2005filling,williams2007terahertz} is impractical due to the need for external lasers, such as quantum cascade lasers or infra-red lasers. These lasers have relatively large apertures, and hence they can only feed a not-so-small number of AEs at the same time. On the other hand, electrical pumping can be achieved via graphene-based sub-micrometric high-electron mobility transistors (HEMTs). Following signal generation by HEMT nano-transceivers, a combination of plasmonic waveguides \cite{Leon6193116} and phase controllers \cite{Singh7481218} delivers these signals to the desired AEs. However, the latter approach is hindered by the low power (a few microwatts) of nano-transceivers \cite{knap2008plasma}, as well as the short  propagation length (a few wavelengths) of SPP waves \cite{akyildiz2017graphene}. Nevertheless, it is argued \cite{akyildiz2016realizing} that since such plasmonic sources are very small, they can be directly attached to each AE, which results in high radiated power. 

\newpage

Attaching plasmonic sources to nano-antennas results in a simplified fully-digital architecture, in which each AE can be individually controlled. This is accounted for in our system model by the precoding and combining matrices, ${W}_t$ and ${W}_r$ of \eqref{eq:channel}, when $Q\!=\!1$. However, the ability to fully control the graphene sheet comes at a hardware cost, since a separate  radio frequency (RF) chain would be required to feed each AE. Nevertheless, with $Q\!>\!1$, the proposed schemes are hybrid SM \cite{He8107553}, in which only one RF chain needs to be activated per SA, which significantly reduces power consumption.

\section{Performance Analysis}
\label{sec:probSER}

An upper bound on the SER performance of SM can be derived using the pairwise error probability (PEP), $P(\mbf{x}\rightarrow \dot{\mbf{x}})$, which is the probability of erroneously deciding on a vector $\dot{\mbf{x}}$ when the true transmitted vector is $\mbf{x}$. We have \cite{5378677Handte,Guo7929326}:
\begin{equation}\label{eq:PEP}
	P(\mbf{x}\rightarrow \dot{\mbf{x}}) = Q\left(  \sqrt{ \frac{ \norm{\mbf{H}\left( \mbf{x} - \dot{\mbf{x}}\right)}^2}{2\sigma^2} }\right).
\end{equation} 
This bound is calculated by summing the PEPs of all possible combinations of symbol vectors and averaging over $\mbf{H}$. Note that for a THz-band LoS environment, we do not assume channel fading of a specific distribution, but rather a deterministic channel. Hence, averaging over $\mbf{H}$ is not required. This bound corresponds to the case of optimal ML detection \cite{5378677Handte}, where antenna indices and symbols are jointly detected by exhaustively searching over all their possible combinations. This is computationally intensive in an UM-MIMO setup. In this work, we only consider MRRC as a detection scheme, in which the antenna index and the transmitted symbol are separately detected. In fact, MRRC acts as an upper bound on the achievable error rate performance of more sophisticated detectors \cite{Maleki7370947}. In the following, we formulate theoretical SER equations that approximate the performance of our proposed solutions. We limit the analysis to the more elaborate second mode of operation (\emph{Region 2}).

Following a similar formulation for SM over Rayleigh fading channels at low frequencies \cite{Mesleh4382913}, we note that the sources of error in MRRC are of two types: erroneously estimating the antenna number (SA or AE), and erroneously detecting the transmitted symbol. We denote by $P_a$ the probability of the former and by $P_s$ the probability of the latter. The overall SER probability $P_e$ can be expressed as 
\begin{equation}\label{eq:SER1}
	P_e = P_a + P_s - P_aP_s.
\end{equation} 
Here again, due to LoS dominance, $\mbf{H}$ is deterministic and no averaging is required to compute $P_s$. The SER of a square $\abs{\mathcal{X}}$-QAM in the presence of AWGN is 
\begin{equation}\label{eq:Ps}
	P_s = 4Q\left( \sqrt{\frac{3\gamma}{\abs{\mathcal{X}}-1}}\right).
\end{equation} 
Let $\mbf{\mu}\!=\![\mu_{1}\mu_{2}\cdots \mu_{\sqrt{\abs{\mathcal{X}}}/2}^{}]\!=\!\Gamma\!\times\!(2\!\times\![1,2,\cdots,\sqrt{\abs{\mathcal{X}}}/2]\!-\!1)$ be the vector of normalized positive real-part (or imaginary-part) components of a QAM constellation, where $\Gamma\!=\!1/(\sqrt{2(\abs{\mathcal{X}}\!-\!1)/3})$. $P_a$ can be expressed as
\begin{align}\label{eq:ber_Pa}
	P_a & = (1-\tilde{P}_a)(1-\tilde{P}_a) = 2\tilde{P}_a - \tilde{P}_a^2 \\
	\tilde{P}_a & = \frac{2}{\sqrt{\abs{\mathcal{X}}}} \sum_{i=1}^{\sqrt{\abs{\mathcal{X}}}/2} P(\mu_i),
\end{align}
where $\tilde{P}_a$ is the average probability of antenna estimation error over $\mbf{\mu}$ (average error when only considering either the real or the imaginary part of the symbols), and $P(\mu_i)$ is the probability of antenna estimation error given a specific $\mu_i$. 

Let $\mbf{h}_k$ be the channel column vector that consists of the channel coefficients from the selected transmitting SA $k$ to all receiving SAs. The effective received vector, assuming identity precoding and combining matrices, is $\mbf{y}=\mbf{h}_k x_k +\mbf{n}$. After MRRC, $\mbf{g}$ can be expressed as
\begin{equation}\label{eq:g_vec}
	\mbf{g} = \left[ \begin{tabular}{c} 
	$\mbf{h}_1^{\mathcal{H}}\mbf{h}_kx_k +\mbf{h}_1^{\mathcal{H}}\mbf{n}$ \\
	 $\mbf{h}_2^{\mathcal{H}}\mbf{h}_kx_k +\mbf{h}_2^{\mathcal{H}}\mbf{n}$ \\
	 \vdots \\ 
	 $\mbf{h}_{M_t\!\times\!N_t}^{\mathcal{H}}\mbf{h}_kx_k +\mbf{h}_{M_t\!\times\!N_t}^{\mathcal{H}}\mbf{n}$ \end{tabular} \right].
\end{equation} 
Since in \emph{Region 2} near-orthogonality is guaranteed, we have
\begin{equation}\label{eq:g_vec2}
	\mathsf{E}[\mbf{h}_l^{\mathcal{H}}\mbf{h}_k] = \left\{
                \begin{array}{c}
                  G_tG_rQ^2\abs{\alpha_{k,k}}^2, \ \text{if} \  l=k  \\
                  \epsilon, \ \ \text{otherwise},
                \end{array}
              \right.
\end{equation} 
where $\epsilon$ is a complex number with a near-zero absolute value. Hence, the distributions of all elements $l$ of $\mbf{g}$ can be approximated as  $\mathcal{N}(\bar{\mu}_{l,i},\dot{\sigma}^2)\!=\!\mathcal{N}(0,\dot{\sigma}^2)$ ($\dot{\sigma}^2$ is a scaled noise variance due to MRRC), except for the $\nth{k}$ element, which behaves as $\mathcal{N}(\bar{\mu}_{k,i},\dot{\sigma}^2)$, where $\bar{\mu}_{l=k,i} \!=\! G_tG_rQ^2\abs{\alpha_{k,k}}^2\mu_i$. 

The probability density function (pdf) of the random variable $v\!=\!\abs{g_l}$ is the folded normal distribution
\begin{equation}\label{eq:fv}
	f_V(v|\bar{\mu}_{l,i},\dot{\sigma}^2) = \frac{1}{\dot{\sigma}\sqrt{2\pi}} \left( e^{-\frac{(v-\bar{\mu}_{l,i})^2}{2\dot{\sigma}^2}} + e^{-\frac{(v+\bar{\mu}_{l,i})^2}{2\dot{\sigma}^2}} \right).
\end{equation} 
Furthermore, computing the estimate SA position $\hat{l}$ \eqref{eq:sym_det} is equivalent to searching for the element with the largest absolute value in a set of $M_t\!\times\!N_t$ elements. Hence, this can be captured by evaluating the pdfs of sorted random variables of pdfs $f_V(v|\bar{\mu}_{l,i},\dot{\sigma}^2)$ and cumulative density functions $F_V(v|\bar{\mu}_{l,i},\dot{\sigma}^2)$, but different means, via order statistics $X_{(M_tN_t)}\!>\!X_{(M_tN_t-1)}\!>\!\cdots\!>\!X_{(1)}$ \cite{casella2002}. The pdf of $X_{(j)}$ is
\begin{align}\label{eq:cont_dist}
	f_{V_{(j)}}(v|\bar{\mu}_{l,i},\dot{\sigma}^2) & = \frac{M_tN_t!}{(j-1)!(M_tN_t-j)!} f_{V}(v|\bar{\mu}_{l,i},\dot{\sigma}^2) \\ 
	& \times \left[ F_{V}(v|\bar{\mu}_{l,i},\dot{\sigma}^2) \right]^{j-1} \\
	& \times \left[ 1 - F_{V}(v|\bar{\mu}_{l,i},\dot{\sigma}^2) \right]^{M_tN_t-j},
\end{align} 
and by analogy with \cite{Mesleh4382913}, we have
\begin{equation}\label{eq:Pui}
	P(\mu_i) = \frac{1}{M_tN_t-1} \left( \sum_{j=1}^{M_tN_t-1} \int_0^{v_t} f_{V_{(M_tN_t)}}(v|\bar{\mu}_{M_tN_t,i},\dot{\sigma}^2) dv  \right),
\end{equation} 
where $v_t$'s denote the intersection points between the plot of $f_{V_{(M_tN_t)}}(v|\bar{\mu}_{M_tN_t,i},\dot{\sigma}^2)$ and the plots of $f_{V_{(j)}}(v|0,\dot{\sigma}^2)$'s. Substituting back in equations \eqref{eq:ber_Pa} and \eqref{eq:SER1}, $P_e$ can be obtained.

\begin{figure}[t]
\centering
\includegraphics[width=3.5in]{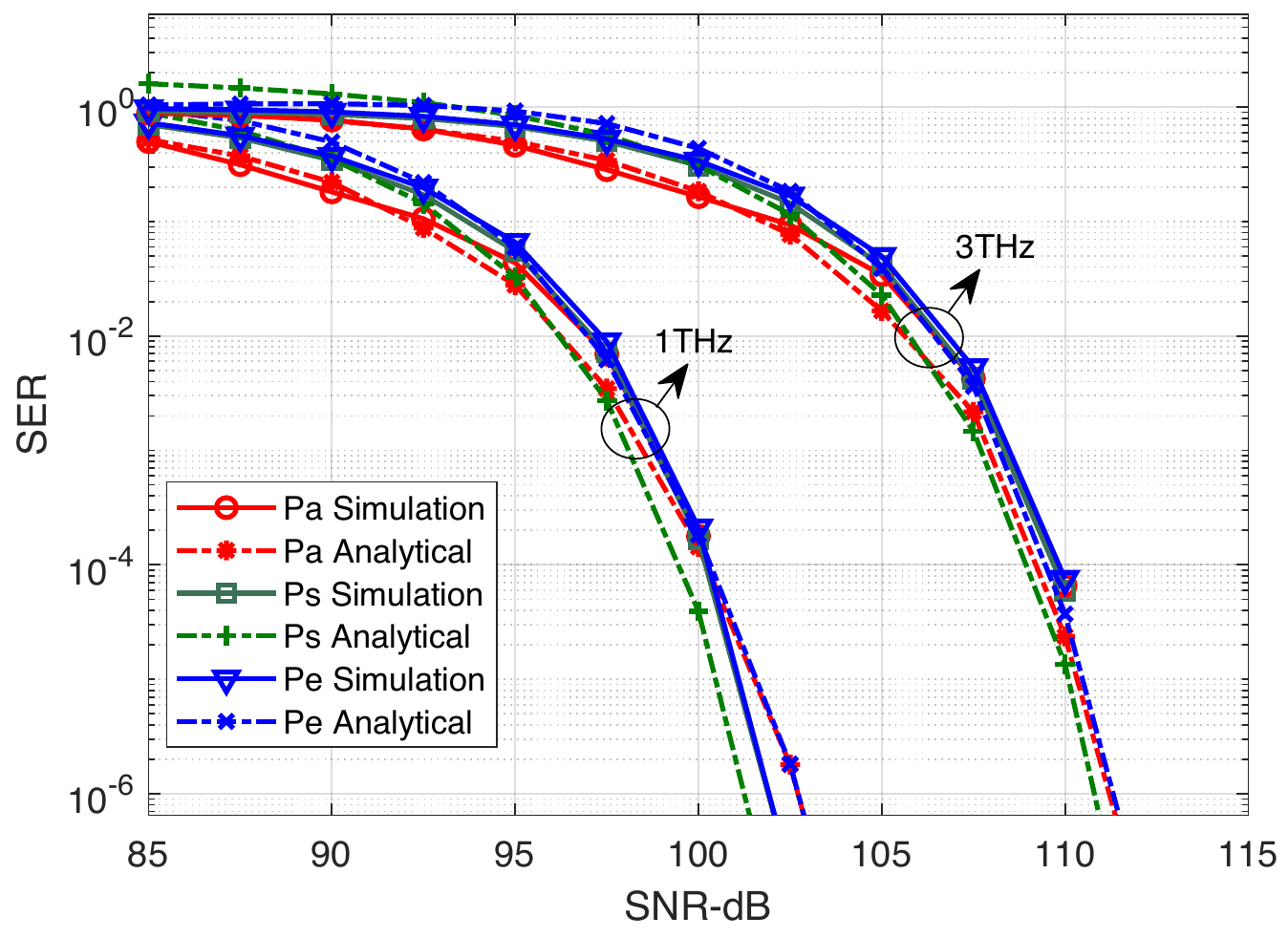}
\caption{Analytical versus simulated SERs: $D\!=\!\unit[1]{m}$, $M\!=\!4$, and $\abs{\mathcal{X}}=16$.}
\label{f:analytical}
\end{figure}

\section{Numerical Results}
\label{sec:simulations}

\begin{figure}[t]
\centering
\includegraphics[width=3.45in]{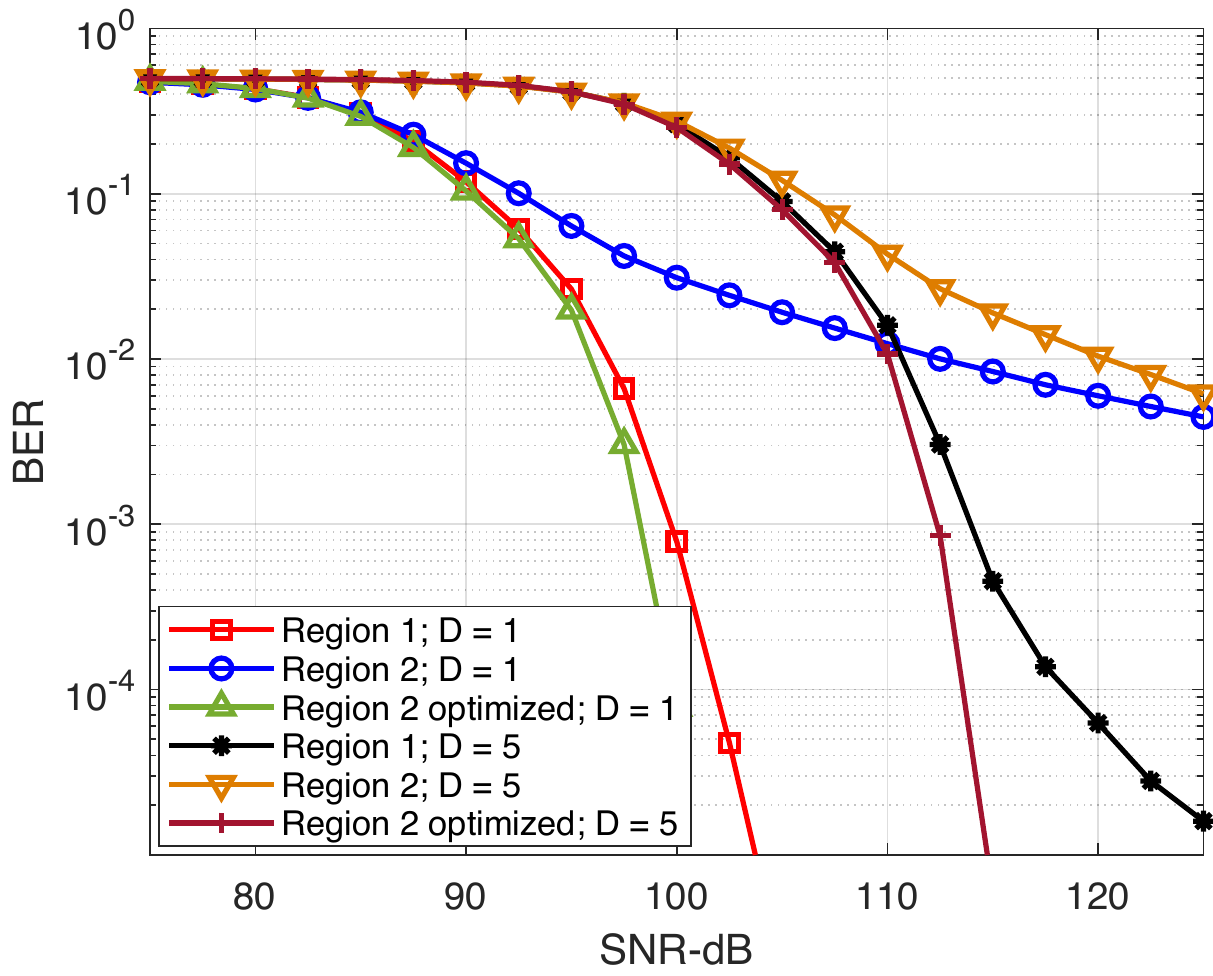}
\caption{BER performance: $f=\unit[1]{THz}$, $M=4$, $Q=1$, and $\abs{\mathcal{X}}=16$.}
\label{f:BER1}
\end{figure}

\begin{figure}[t]
\centering
\includegraphics[width=3.3in]{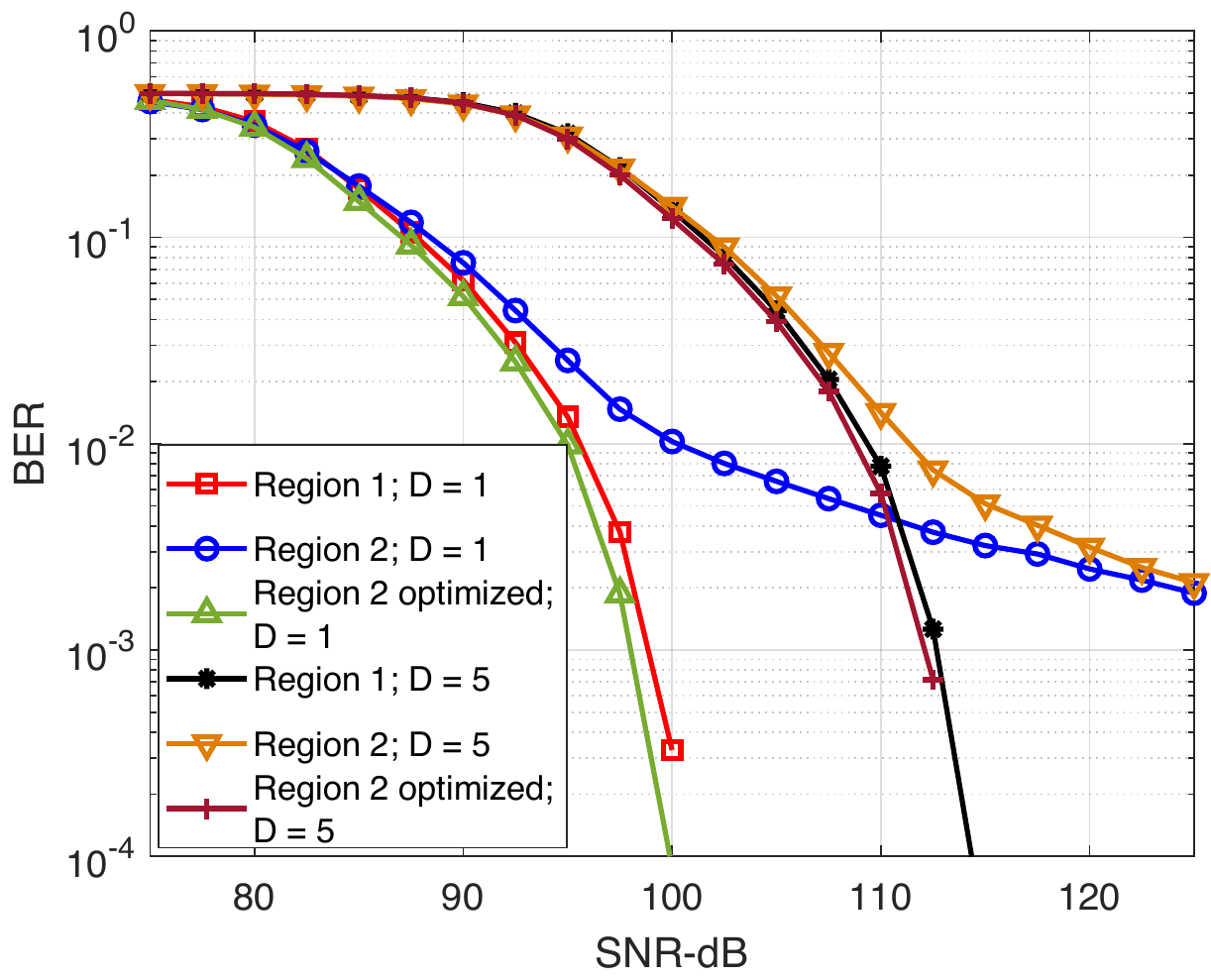}
\caption{BER performance: $f=\unit[1]{THz}$, $M=8$, $Q=1$, and $\abs{\mathcal{X}}=64$.}
\label{f:BER2}
\end{figure}


\begin{figure}[t]
\centering
\includegraphics[width=3.4in]{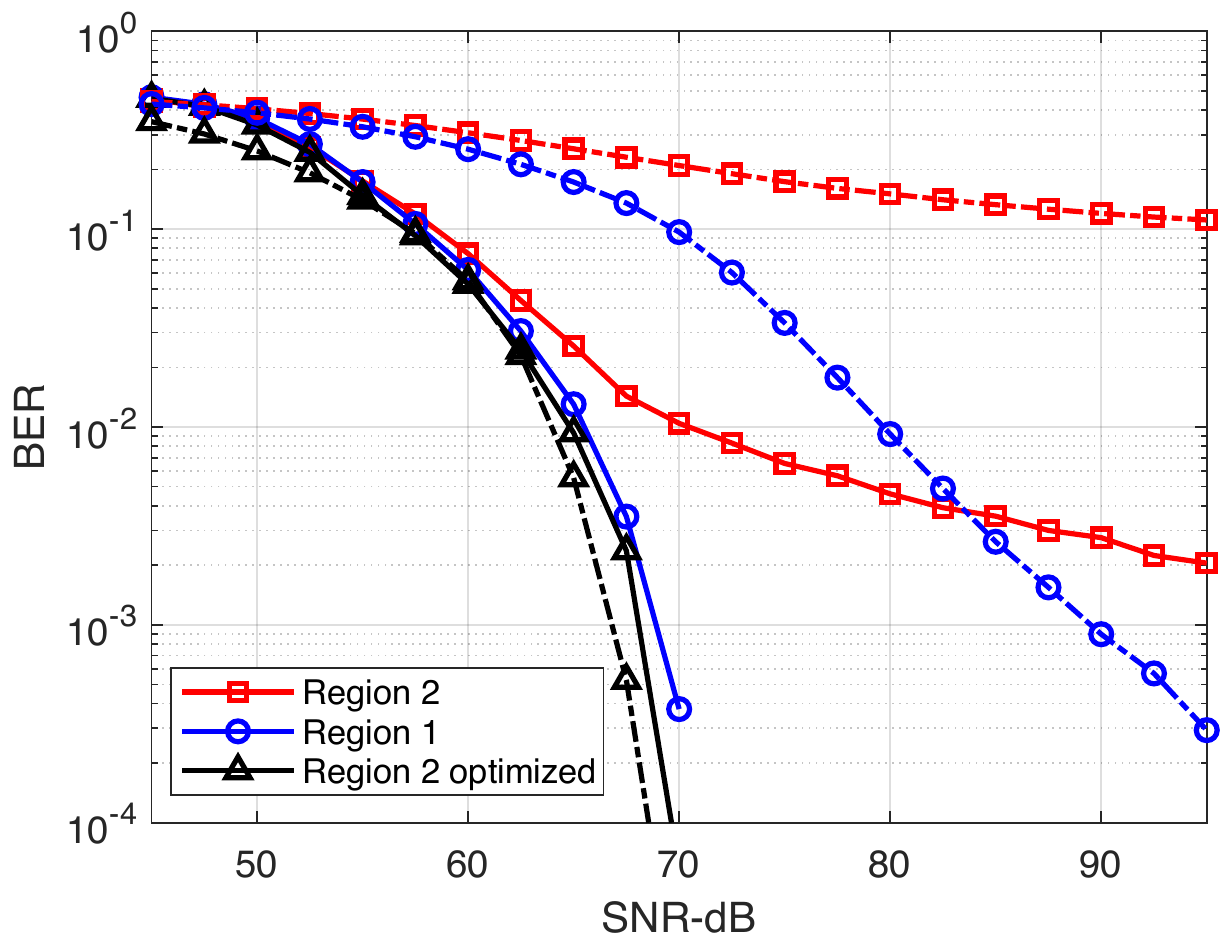}
\caption{BER performance of SM (solid lines) and SMX (dotted lines): $f=\unit[1]{THz}$, $D=\unit[1]{m}$,    $M=8$, $Q=32$ and $\abs{\mathcal{X}}=64$.}
\label{f:BER4}
\end{figure}

We simulated the proposed SM schemes following the system model of Sec.~\ref{sec:sysmodel}. Theoretical versus simulated SER plots are shown in Fig. \ref{f:analytical}, for a $16\!\times \!16$ MIMO system ($M\!=\!4$) with 16-QAM, at $\unit[1]{THz}$ and $\unit[3]{THz}$, and for $D\!=\!\unit[1]{m}$. The second mode of operation was assumed, with $\Delta$ tuning as per \eqref{eq:innerprod}. The simulation and analytical results are in close agreement for all components of error probability, $P_e $, $P_a$, and $P_s$. It is shown that increasing the frequency of operation from $\unit[1]{THz}$ to $\unit[3]{THz}$ results in a $\unit[10]{dB}$ loss in SNR. Note that no beamforming was assumed for these plots. Hence, a combination of array gains, transmit antenna gains and receive antenna gains can shift the high SNR values down.


The relative bit error rate (BER) performance of the proposed schemes is illustrated in figures \ref{f:BER1} and \ref{f:BER2}, with different communication ranges at $\unit[1]{THz}$, for $16\!\times \!16$ MIMO ($M\!=\!4$) with $64$-QAM and $64\!\times \!64$ MIMO ($M\!=\!8$) with $64$-QAM, respectively. While the first mode of operation (\emph{Region 1}) guarantees good performance due to average good channel conditions,  the second mode (\emph{Region 2 optimized}: operations in \emph{Region 2} with proper tuning of $\Delta$) achieves the best performance. Furthermore, both modes significantly outperform the case of operating in \emph{Region 2} without optimizing antenna spacing (risk of operating near the yellow curves of Fig. \ref{f:channel_2}). The performance of the latter becomes severely distorted and diverges at high SNR, which emphasizes the importance of our proposed solutions. We noted the same relative performances at $D\!=\!\unit[1]{m}$ and $D\!=\!\unit[5]{m}$, with an SNR gap between the two that is similar to that between operations at $\unit[1]{THz}$ and $\unit[3]{THz}$.

Finally, Fig.~\ref{f:BER4} shows the results for a $64\times 64$ MIMO system when $1024$-element SAs are assumed ($Q\!=\!32$), which introduce a $\unit[30]{dB}$ array gain. We note the same relative performance but at a much lower SNR range. This illustrates the importance of UM-MIMO at the THz-band. Alongside SM, Fig.~\ref{f:BER4} illustrates reference SMX BER curves. Note that antenna-spacing optimization similarly enhances the performance of SMX. In fact, SMX is more sensitive to channel conditions than SM. While SMX slightly outperforms SM under optimal channel conditions, its performance is severely deteriorated otherwise. This further promotes SM as an efficient THz-band paradigm.


\section{Extensions}
\label{sec:extensions}

Having introduced and analyzed THz-band SM, we next present possible future research directions that can overcome the limitations of our proposed solutions, and that can pave the way for efficient utilization of the spatial degrees of freedom at very high frequencies.

\subsection{Frequency-Interleaved Antenna Maps}
\label{sec:interleaved_map}

The distance between two graphene-based plasmonic nano-AEs can be set to $\delta\!=\!\lambda_{\spp}$, which is much smaller than that in metallic antennas, while still avoiding mutual coupling effects \cite{Zakrajsek7928818}. This is due to a confinement factor 
\begin{equation}\label{eq:confinement}
	\eta = \frac{\lambda}{\lambda_{\spp}} \gg 1.
\end{equation} 
In particular, $\eta\!=\!25$ for graphene \cite{Abadal7047773}. SPP waves are electromagnetic waves that propagate in the interface between a metal and a dielectric, due to global oscillations of electric charges, at a much lower speed compared to electromagnetic waves in free space; hence the confinement factor. 

\begin{figure}[t]
\centering
\includegraphics[width=3in]{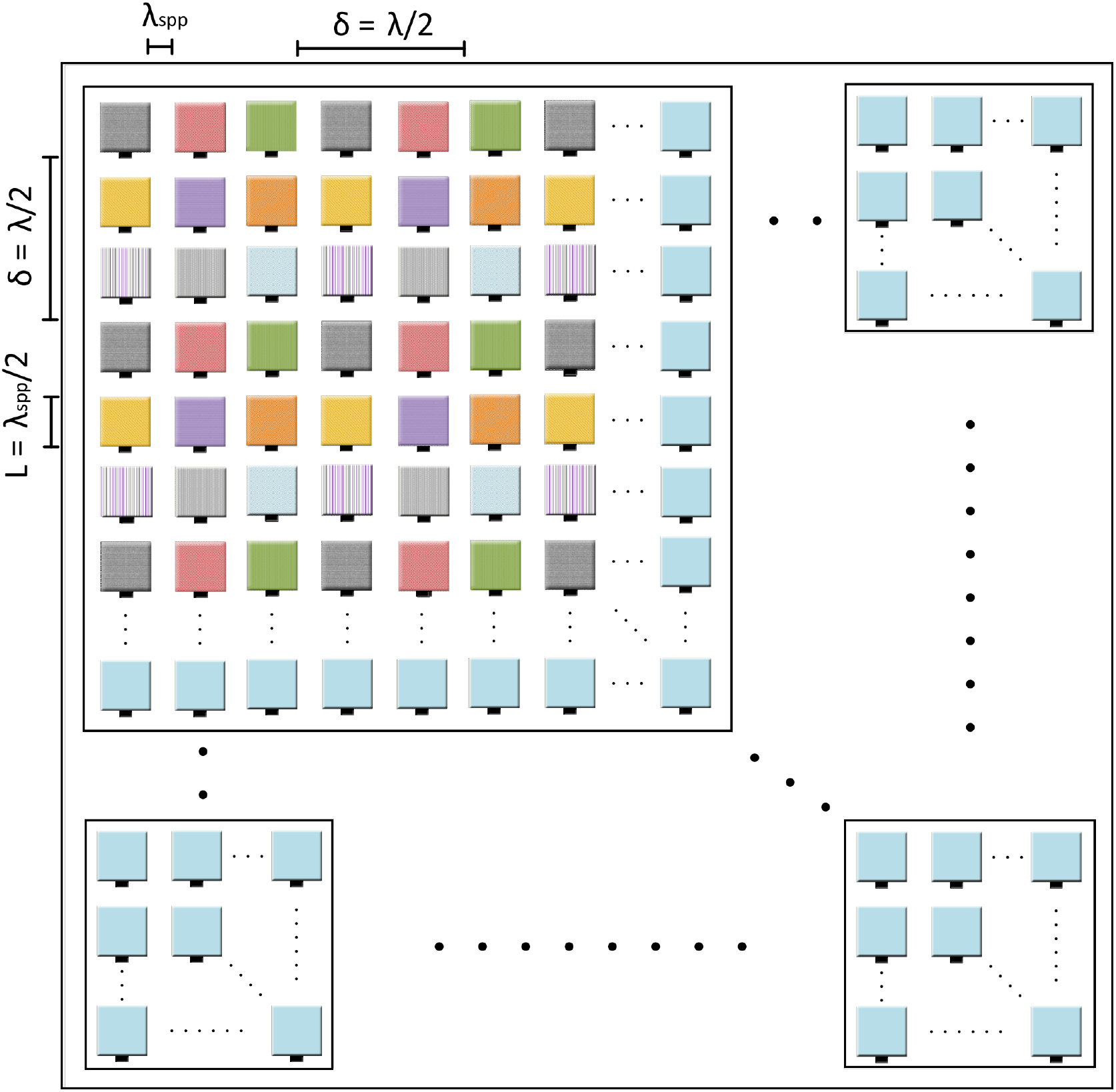}
\caption{Frequency-interleaved AoSA design at the level of AEs.}
\label{f:interleaved_2}
\end{figure}

While such confinement results in favorably dense packaging, it also results in limited beamforming gains due to inadequate spatial sampling. In order to maximize the beamforming gain and minimize side lobes, a rule of thump is to set $\delta\!=\!\lambda/2\!\gg\!\lambda_{\spp}$. Hence, to maintain a compact design while respecting minimal separations, a solution is proposed \cite{zakrajsek2017design} in which non-contiguous nano-AEs are tuned to identical frequencies, while nano-antennas at different frequencies are interleaved in between, along both axes. Similarly, much larger same-frequency SA separations $\Delta$ are required to achieve good multiplexing or SM gains. Therefore, a sparse interleaved antenna map that can accommodate a wider range of operating frequencies is required.  

\begin{figure}[t]
\centering
\includegraphics[width=3in]{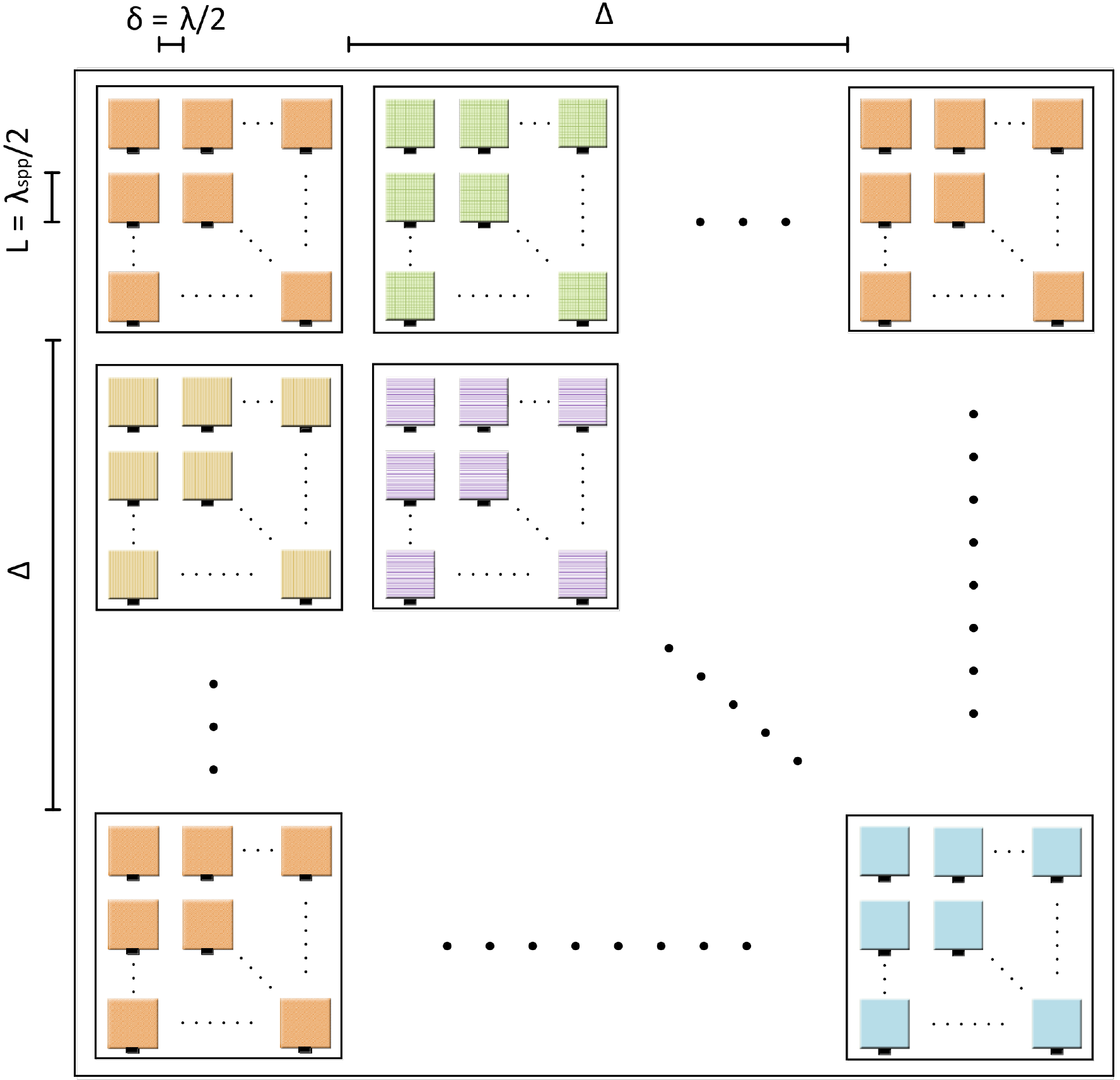}
\caption{Frequency-interleaved AoSA design at the level of SAs.}
\label{f:interleaved_1}
\end{figure}

Illustrations of frequency-interleaved AoSAs are shown in figures \ref{f:interleaved_2} and \ref{f:interleaved_1}, in which interleaving is performed at the level of AEs and SAs, respectively. Note that these schematics are not drawn to scale. In Fig. \ref{f:interleaved_2}, the effective same-frequency inter-AE separation is retained at $\delta\!=\!\lambda/2$, and multiple AEs are fit in between. In Fig. \ref{f:interleaved_1}, same-frequency AEs are also set $\lambda/2$ apart, to achieve good beamforming gains, but same frequency SAs are activated at a sufficiently large $\Delta$ as per our optimization. Nevertheless, these two maps can be combined, where interleaving can be introduced at both levels, but at the expense of higher complexity. Noting that the characteristic length of each AE is $L\!=\!\lambda_{\spp}/2$ \cite{zakrajsek2017design}, the footprint of the active  part of the graphene sheet can be computed as 
\begin{equation}\label{eq:footpring}
	F = \left(\frac{3}{2}MQ\lambda_{\spp}\right)^2.
\end{equation}

The key enabler for such frequency-interleaved maps is the ability to dynamically tune each AE to a specific resonant frequency without modifying its physical dimensions. This can be achieved by modifying the conductivity of these elements, which is a function of the Fermi energy of graphene. Simple material doping can thus modify the Fermi energy and tune an AE. Electrostatic bias is also a viable alternative. However, at relatively lower frequencies, say $f\!<\!\unit[1]{THz}$, such tunability is not easily realized because of the limitations of plasmonic antennas. Therefore, software-defined plasmonic metamaterials \cite{liaskos2015design} have been proposed as an alternative solution.


\subsection{Relaxing Ideal Assumptions}
\label{sec:non-ideal}

Due to severe propagation properties and high beamforming gains at the THz-band, we only assume a LoS component in our system model. However, to be more precise, we can account for a few additional reflected rays. The frequency-dependent NLoS path gain can be expressed as

\begin{align}\label{eq:NLoS}
	\alpha_{m_rn_r,m_tn_t}^{\NLoS} = & \frac{c}{4\pi f (r_{m_rn_r,m_tn_t}^1 + r_{m_rn_r,m_tn_t}^2)} \\& \times  e^{ -\frac{1}{2} \mathcal{K}(f) \left(r_{m_rn_r,m_tn_t}^1 + r_{m_rn_r,m_tn_t}^2\right)}  \\& \times R(f) e^{ -j \frac{2\pi f}{c} \left(r_{m_rn_r,m_tn_t}^1 + r_{m_rn_r,m_tn_t}^2\right)},
\end{align} 
where $r_{m_rn_r,m_tn_t}^1$ and $r_{m_rn_r,m_tn_t}^2$ are the distances between the reflector and the transmitter and the reflector and the receiver, respectively. The value of the path loss threshold $PL_{\thh}$ in equation \eqref{eq:link_budget} decreases by taking advantage of multipath components. Moreover, equation \eqref{eq:sysmodel} can be modified accordingly by employing Rician fading \cite{paulraj2003introduction}.

Several other assumptions have been made in this paper, which can be relaxed in future works. For example, we can account for channel correlation by using the Kronecker model \cite{Forenza1388958,Koca6399330}. Also, we can propose novel models to approximate RF  impairments, which are not yet very well understood for THz-band tranceivers. More importantly, imperfect channel estimation can be taken into consideration. In fact, obtaining channel state information at very high frequencies is very challenging. A solution \cite{Gao7582545} is proposed in this regard in the context of THz beamspace massive MIMO, in which fast channel tracking mechanisms are applied. Such solutions can be combined with other findings on channel estimation for SM \cite{Wu6945322,Sugiura6319351}. Furthermore, the effect of blockage at high frequencies should not be neglected. In fact, blockage can generate over the medium due to obstacles that block narrow beams, or at the source due to suspended particles at the antenna arrays that can block small AEs \cite{Eltayeb8248776}.  

Another stringent assumption of this work is assuming carrier-based modulation. While continuous transmission is supported by plasmonic antennas at the low-THz post-mmWave range, it is still very difficult to generate more than short pulses with graphene at room temperature. Time-domain pulse-based single-carrier modulation can thus be used, with a frequency response over the THz band, and with a corresponding power in the order of a few milli-watts. Since this power is not sufficient for long-distance communications, nano-scale applications are typical candidates for pulse-based modulation. To this end, pulse-based asymmetric on-off keying modulation spread in time (TS-OOK) is proposed in \cite{Jornet6804405}, which consists of trading very short pulses (one hundred femtosecond-long) among nano-devices as a logic 1. This scheme supports a very large number of nano-devices that can transmit at very high rates, ranging from few Gb/sec to few Tb/sec. 

Note that adaptive systems at the THz-band can change modulation types depending on system and channel conditions. An interesting signal processing exercise would then be to blindly estimate these modulations at the receiver side \cite{8540324Iqbal}, in an extension to the well-investigated modulation classification problem. Furthermore, the proposed solutions of this work can be adapted to pulse-based modulations by borrowing results from optical SM \cite{Mesleh5722087}. However, judging by the pace of recent advancements in graphene-based technologies, it is expected that continuous modulation schemes over the high-THz spectrum will be supported in the not-too-distant future.

\subsection{Generalized Spatial and Index Modulation}
\label{sec:ext_GSM}

A straightforward extension of this work is to enable the selection of a specific combination of antennas at a specific instance, which would in effect be a form of generalized SM (GSM) \cite{Younis5757786,Datta6554991}. In fact, GSM is particularly of interest at high frequencies due to the large number of AEs in small footprints. For an AoSA of our system model, the total number of transmitted bits in one channel use would be
\begin{equation}\label{eq:nbofbitsGSM}
	N_b^{(\GSM)} =  \log_2 \left \lfloor {S \choose M^2Q^2} \right \rfloor + \log_2\left( \abs{\mathcal{X}} \right),
\end{equation}  
where $S$ is the number of AEs that can be simultaneously activated. The literature is rich in SM and GSM solutions for massive MIMO \cite{Basnayaka7248610,Zuo8171123}. However, even more efficient low-complexity algorithms are required at such large scales. 

Furthermore, the frequency dimension can be exploited by assigning information bits to the interleaved maps of Sec. \ref{sec:interleaved_map}, to further increase the spectral efficiency. For example, the same AE can be assigned multiple frequencies at different uses, where each specific frequency is mapped to specific information bits. This approach is particularly efficient at the THz-band due to the large number of available spectral windows that are created by the absence of molecular absorption (Fig. \ref{f:channel_1}). The resultant scheme can be referred to as generalized index modulation (GIM), in which
\begin{equation}\label{eq:nbofbitsGIM}
	N_b^{(\GIM)} =  \log_2 \left \lfloor {\bar{F} \choose F} \right \rfloor+ \log_2 \left \lfloor {S \choose M^2Q^2} \right \rfloor + \log_2\left( \abs{\mathcal{X}} \right),
\end{equation}  
where $F$ is the number of available narrow frequency bands, and $\bar{F}$ is the number of frequency bands that can be simultaneously utilized. Note that $N_b^{(\GIM)}$ can be further increased if the frequency and spatial maps are jointly realized. The resultant scheme would guarantee best utilization of the available spectrum at the THz-band. Note, however, that the applicability of GIM is limited by the speed of frequency transitions, which is determined by the speed of change in Fermi energy, through material doping or electrostatic bias (fast frequency modulation rates are expected in future technologies). Furthermore, detecting information symbols from a huge number of possible combinations is challenging. A possible solution is to apply artificial intelligence and machine-learning techniques to determine the indices at the receiver.

\subsection{Enhanced Detection and Coding Schemes}
\label{sec:ext_detection}

A plethora of SM detection algorithms can be studied in the context of THz communications that can trade performance with computational complexity. One possible detector can make use of the inherent hierarchy in the studied AoSA. For example, in the first mode of operation (Sec. \ref{sec:dual_layer}), the SA index can be detected first, followed by detecting the AE index. Hence, MRRC can be used to detect the former, while ML detection treats the latter jointly with the transmitted symbol. This is motivated by the fact that having $\Delta\!>\!\delta$ means that errors are more likely to occur at the AE level. The same concept can be extended to adaptive channel coding. Bits corresponding to AE indices can be protected by a stronger coding scheme, with more redundancy compared to the coding scheme that protects bits corresponding to SA indices.  

Furthermore, in a GSM setup, more elaborate detection schemes should be considered, which can achieve near-optimal performance with reasonable complexity. Typical candidates are variations of large MIMO detection schemes. Recently proposed large MIMO detectors are based on lattice reduction~\cite{Zhou_LR_2013}, message passing on graphical models \cite{Narasimhan_2014}, Monte Carlo sampling \cite{Datta_2013}, local search criteria \cite{Li_2010}, heuristic tabu search algorithms \cite{Srinidhi_2011},  and sub-space decompositions \cite{Sarieddeen8186206}.

\subsection{Optimization Problems}
\label{sec:ext_opt}

Antenna array design and resource allocation criteria can be formulated as optimization problems. Since the number of parameters that can be optimized is very large, such optimizations can be carried over selected sub-problem formulations. For example, in \cite{zakrajsek2017design}, optimal allocation of frequencies to AEs is formulated, with the objective of maximizing capacity subject to the number of AEs, the number of available frequencies, antenna gains, array factors, beamsteering angles, and absorption coefficients. In other studies \cite{Han7490372,Han7321055}, the pulse waveform design is optimized by adapting the number of frames and power allocation to combat the losses over a specific communication distance. Furthermore, beamforming at the THz-band can be optimized, as in \cite{Lin7116524}, where an adaptive hybrid beamforming scheme is proposed that is also distance-aware. Such optimization problems can be modified and appended with SM-specific constraints.

Finally, antenna array design optimization can be extended beyond the scope of this work, in which we assumed that the proposed graphene sheets are sufficiently large such that $\bar{\Delta}$ can always be secured. In fact, due to limited physical sizes at the transmitter and the receiver, near-orthogoal channels, and hence high multiplexing gains, can not be supported beyond a so-called \emph{Rayleigh distance}. Point-to-point LoS communication beyond this distance has been studied for mmWave communications, for both uniform \cite{6800118Wang} and non-uniform \cite{7546944Wang} linear arrays. These studies can be extended in the context of configurable THz-band antennas. Nevertheless, spatial oversampling can be considered, in which AE separations can be decreased below $\lambda_{\spp}$. This oversampling is argued to decrease the spatio-temporal frequency-domain region of support of plane waves \cite{8732419Rappaport}, which results in multiple benefits such as reduced noise figures and increased linearity.

\section{Conclusion}
\label{sec:conclusion}

In this paper, the applicability of SM to THz-band communications have been studied. We first addressed the peculiarities of the THz channel by deriving the optimal antenna spacings that result in favorable propagation conditions. Following that, we proposed two SM modes of operation for an AoSA architecture of graphene-based antenna arrays. We derived analytical closed-form SER equations to give further insight on the performance of the proposed solutions, and we demonstrated through empirical simulations that optimizing active AE positions results in significant performance enhancement. We concluded the paper by proposing several future research directions that could help in realizing THz-band SM. 




\newpage

\begin{IEEEbiography}[{\includegraphics[width=1in,height=1.25in,clip,keepaspectratio]{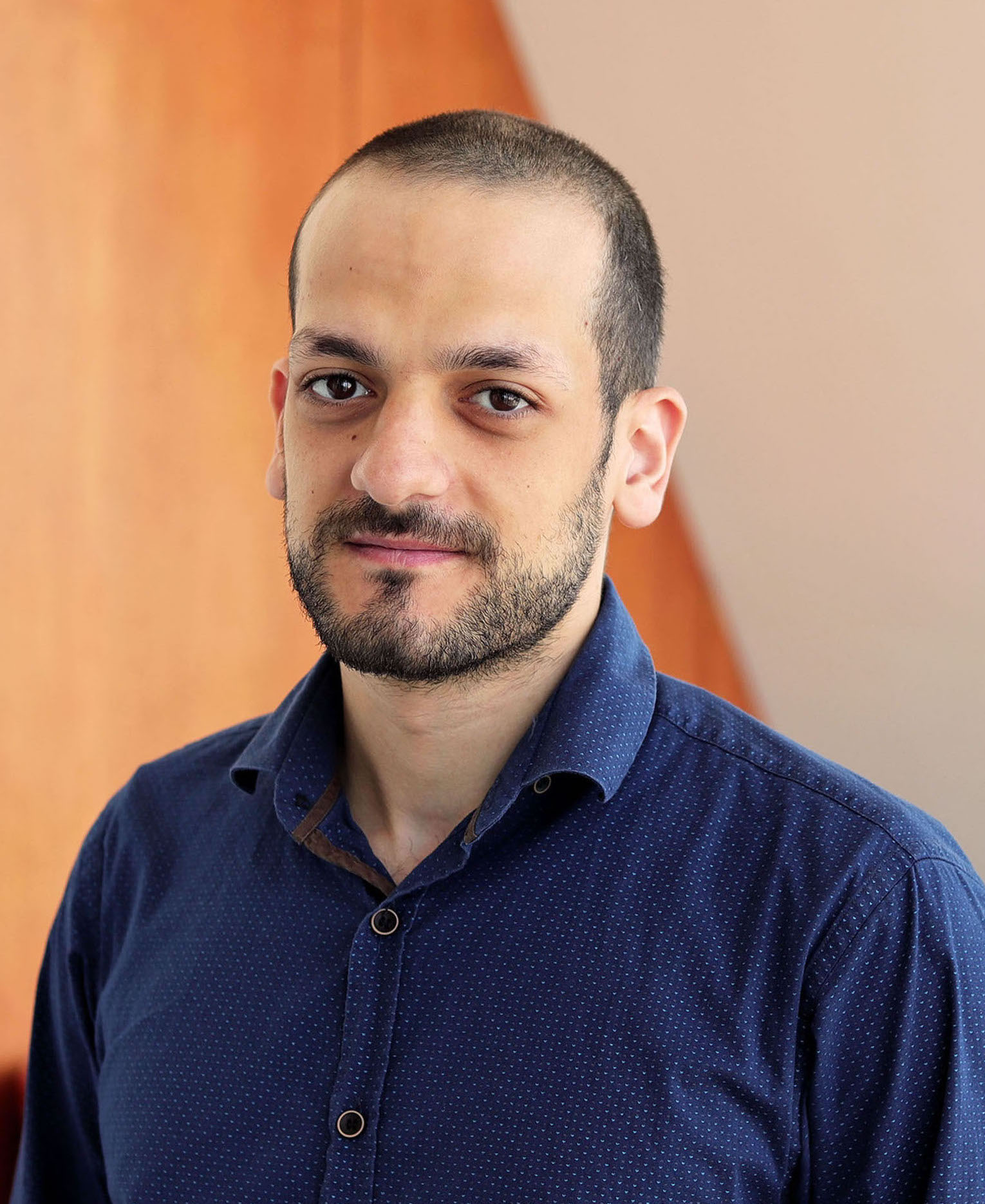}}]
{Hadi Sarieddeen} (S'13-M'18) received the B.E. degree (summa
cum laude; first in graduating class) in computer and communications engineering from Notre Dame University-Louaize (NDU), Zouk Mosbeh, Lebanon, in 2013, and the Ph.D. degree in electrical and computer engineering from the American University of Beirut (AUB), Beirut, Lebanon, in 2018.
He is currently a postdoctoral research fellow in the Computer, Electrical and Mathematical Sciences and Engineering (CEMSE) Division at King
Abdullah University of Science and Technology (KAUST), Thuwal, Makkah Province, Saudi Arabia. His research interests are in the areas of communication theory and signal processing for wireless communications, with emphasis on large, massive, and ultra-massive MIMO systems and terahertz communications.
\end{IEEEbiography}

\begin{IEEEbiography}[{\includegraphics[width=1in,height=1.25in,clip,keepaspectratio]{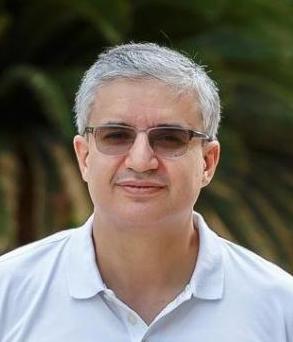}}]{Mohamed-Slim Alouini} (S'94-M'98-SM'03-F'09)  was born in Tunis, Tunisia. He received the Ph.D. degree in Electrical Engineering from the California Institute of Technology (Caltech), Pasadena, CA, USA, in 1998. He served as a faculty member in the University of Minnesota, Minneapolis, MN, USA, then in the Texas A\&M University at Qatar, Education City, Doha, Qatar before joining King Abdullah University of Science and Technology (KAUST), Thuwal, Makkah Province, Saudi Arabia as a Professor of Electrical Engineering in 2009. His current research interests include the modeling, design, and performance analysis of wireless communication systems.
\end{IEEEbiography}

\begin{IEEEbiography}[{\includegraphics[width=1in,height=1.25in,clip,keepaspectratio]{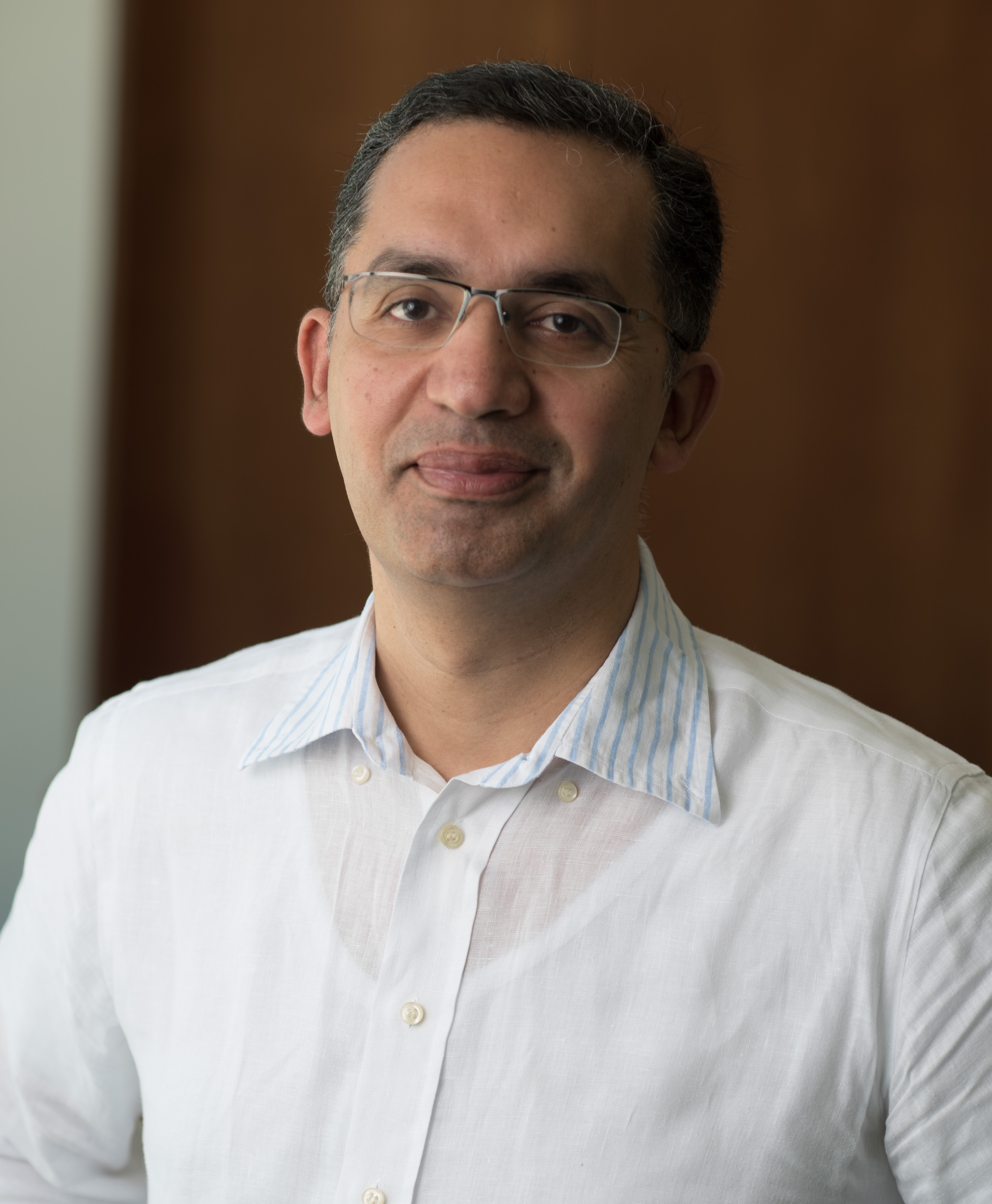}}]{Tareq Al-Naffouri} (M'10-SM'19) received the B.S. degrees in mathematics and electrical engineering (with first honors) from King Fahd University of Petroleum and Minerals, Dhahran, Saudi Arabia, the M.S. degree in electrical engineering from the Georgia Institute of Technology, Atlanta, in 1998, and the Ph.D. degree in electrical engineering from Stanford University, Stanford, CA, in 2004.

He was a visiting scholar at California Institute of Technology, Pasadena, CA in 2005 and summer 2006. He was a Fulbright scholar at the University of Southern California in 2008. He has held internship positions at NEC Research Labs, Tokyo, Japan, in 1998, Adaptive Systems Lab, University of California at Los Angeles in 1999, National Semiconductor, Santa Clara, CA, in 2001 and 2002, and Beceem Communications Santa Clara, CA, in 2004. He is currently an Associate Professor at the Electrical Engineering Department, King Abdullah University of Science and Technology (KAUST). His research interests lie in the areas of sparse, adaptive, and statistical signal processing and their applications, localization, machine learning, and network information theory. He has over 240 publications in journal and conference proceedings, 9 standard contributions, 14 issued patents, and 8 pending.

Dr. Al-Naffouri is the recipient of the IEEE Education Society Chapter Achievement Award in 2008 and Al-Marai Award for innovative research in communication in 2009. Dr. Al-Naffouri was an Associate Editor of IEEE Transactions on Signal Processing 2013-2018.

\end{IEEEbiography}

\end{document}